\documentclass{aa}

\usepackage{graphicx}
\usepackage{txfonts}
\usepackage{threeparttable}
\usepackage{orcidlink}

\begin{document} 

   \title{Machine-learning clustering of close-in exoplanet populations: links to pebble accretion}

   \author{Yi Duann
          \orcidlink{0000-0002-4260-385X}
          \inst{1,3}\fnmsep\thanks{Corresponding Author: yiduann@gm.astro.ncu.edu.tw}
          \and
          Anders Johansen
          \orcidlink{0000-0002-5893-6165}
          \inst{1,2}
          \and
          Haiyang S. Wang
          \orcidlink{0000-0001-8618-3343}
          \inst{1}
          \and
          H. Jens Hoeijmakers
          \orcidlink{0000-0001-8981-6759}
          \inst{2}
          }
    \titlerunning{Machine-learning clustering of close-in exoplanet populations: links to pebble accretion}
    \authorrunning{Y. Duann et al.}

   \institute{Center for Star and Planet Formation, Globe Institute, University of Copenhagen, Copenhagen, Denmark
        \and
        Lund Observatory, Department of Physics, Lund University, Lund, Sweden
        \and
        Graduate Institute of Astronomy, National Central University, Taoyuan, Taiwan
             }

\abstract
{Close-in exoplanets span a wide range of orbital architectures and physical properties, consistent with the combined influence of migration processes and formation conditions. Although population synthesis models predict the emergence of distinct planetary populations, establishing a statistically robust connection between observed samples and synthetic populations remains challenging, particularly in high-dimensional parameter spaces dominated by dynamical quantities.}
{Intrinsic population-level organisation within the observed sample of close-in exoplanets is investigated through physically motivated parameters, with the aim of establishing a quantitative link to formation pathways predicted by pebble-accretion models.}
{A two-stage Gaussian mixture model (GMM) is applied to an observed sample of close-in exoplanets, with unsupervised probabilistic clustering performed in a feature space dominated by dynamical descriptors of planet–star interactions. The resulting clusters are mapped onto a pebble-accreted synthetic population within a statistically motivated 3D parameter space. Formation-related quantities, including formation timing, gas fraction, and ice--rock mass ratio, are then employed to establish a probabilistic interpretation and comparison of the mapped clusters.}
{Statistically supported sub-populations are identified without the imposition of predefined classification boundaries, including very-massive gas giants, hot giants, warm-Jupiter-dominated systems, and lower-mass giants. When mapped onto pebble-accreted synthetic populations, these observational clusters exhibit systematic differences in formation timing, gas fraction, and solid growth histories across the dominant populations.}
{Data-driven clustering in a dynamical-parameter space enables a direct and statistically robust comparison between observed exoplanet sub-populations and pebble-accretion synthetic populations. The inferred cluster-level trends point to systematic differences in formation timing and accretion histories among giant populations, with very-massive gas giants preferentially associated with earlier formation epochs than hot-giant and warm-Jupiter-dominated systems. This demonstrates the potential of physically motivated machine-learning approaches to connect observed exoplanets with theoretical formation scenarios at the population level.}


   \maketitle

\section{Introduction}\label{sec1}

The growing census of close-in giant exoplanets exhibits substantial diversity in orbital architectures, bulk densities, and atmospheric properties, indicating that multiple formation and evolutionary pathways are likely at play~\citep{Mayor1995,Winn2015,Dawson2018}. In particular, the relative importance of disk-driven migration and high-eccentricity migration in shaping the close-in gas-giant population remains actively debated, with constraints drawn from population-level trends, system architectures, and detailed studies of individual systems~\citep{Ford2014,Dawson2018}.

Close-in giant planets on low-eccentricity and low-obliquity orbits are commonly interpreted as consistent with disk-driven migration, whereas dynamically excited and misaligned systems are often associated with high-eccentricity migration pathways~\citep{Gupta2024}. Demographic studies further suggest that the occurrence rate of hot Jupiters (HJs) declines with stellar age, potentially reflecting long-term tidal evolution~\citep{Chen2023}. Alternative scenarios such as inside-out planet formation have also been proposed, in which planets form sequentially near pressure maxima associated with the dead-zone inner boundary of protoplanetary disks~\citep{Chatterjee2014,Chatterjee2015,Hu2016,Tan2016,Guillot2023}.

In parallel, machine-learning techniques have become increasingly important tools for analysing large exoplanet catalogues and related stellar datasets~\citep{Alibert2019a,Alibert2019b,Lalande2024,Zhao2024,Alibert2025,Sahlmann2025,Hossain2026}. Supervised approaches are widely used for separating planetary from non-planetary transit signals and refining candidate vetting and validation~\citep{Schanche2019,Sturrock2019,Karimi2025}, while unsupervised and semi-supervised methods have been applied to explore population-level organisation in high-dimensional parameter spaces. These include classifying variable stars, optimising exoplanet atmosphere retrieval, identifying habitable-planet candidates via anomaly detection, and clustering confirmed planets in mass--radius--period space without reliance on predefined labels~\citep{Armstrong2015,Hayes2020,Sarkar2021,MousaviSadr2023,Fotopoulou2024}. Collectively, these studies demonstrate that data-driven approaches can robustly identify intrinsic trends and sub-populations in observed stellar and exoplanet samples, providing an empirically grounded interface to physically motivated models of planet formation and evolution~\citep{Winter2020}.

On the theoretical side, pebble-accretion models provide a well-established framework for the rapid assembly of massive planetary cores and subsequent gas-envelope growth in protoplanetary disks, as reviewed by~\citet{Johansen2017}. Efficient accretion of small solids, particularly beyond the ice line, allows cores to reach critical masses on timescales compatible with disk lifetimes~\citep{Lambrechts2014,Johansen2019}. When combined with gas accretion and migration, pebble-accretion models naturally predict a broad diversity of planetary masses, compositions, and orbital architectures, including multiple populations of giant planets arising from different disk conditions and evolutionary pathways~\citep{Bitsch2015,Bitsch2018,Ndugu2018,Izidoro2021}. In parallel, complementary predictions are provided by global population-synthesis frameworks such as the Bern global planet-formation model, in which planet growth proceeds through planetesimal-driven core accretion, with migration and long-term evolution treated self-consistently~\citep{Alibert2025}.

Despite substantial progress in both observational surveys and planet formation modelling, establishing a direct and statistically robust connection between observed exoplanet populations and synthetic populations produced by formation models remains challenging. Existing comparisons are typically carried out in low-dimensional projections of the data, such as period--radius or irradiation--radius planes, or through calibration of model parameters to reproduce selected demographic features~\citep{Mordasini2012,Fulton2017,Emsenhuber2021}. Data-driven clustering has also been applied to synthetic populations generated by global core-accretion frameworks such as the Bern model, where simulated observables are linked to formation histories within a planetesimal-based, rather than pebble-accretion, growth framework~\citep{Schlecker2021}. In that context, the clustering is performed on the synthetic populations themselves and is not extended to an unsupervised, high-dimensional mapping between observed exoplanet samples and synthetic formation models. 

More recently, machine-learning surrogates have been introduced to bridge this gap more directly. For instance, \citet{Burn2025} trained conditional invertible neural networks on global planet-formation simulations to infer disk-parameter posteriors from planetary mass and semi-major axis. While this demonstrates the feasibility of probabilistic inversion from observables to formation parameters, the results remain strongly dependent on the assumed physical model and the coverage of the synthetic training set, with sparse sampling prone to biased or overconfident extrapolation. This motivates an alternative strategy in which physically interpretable patterns are first identified directly within the observed population, before linking those statistically supported groupings back to formation models. In particular, features that encode fundamental dynamical scales offer a natural starting point for such a data-driven yet physically grounded classification.

Dynamical quantities that encode interaction strengths and characteristic stability scales, such as the Hill radius ($R_\mathrm{Hill}$) and the Safronov number ($\Theta_\mathrm{saf}$), provide physically motivated descriptors of planetary systems~\citep{Safronov1967,Goldreich1982}. While $\Theta_\mathrm{saf}$ has been used to motivate empirical classifications of hot Jupiters through heuristic boundaries in reduced parameter spaces~\citep{Ozturk2018}, and $R_\mathrm{Hill}$ has served as a diagnostic quantity in studies of orbital stability, resonant configurations, and migration~\citep{Wu2019,Tamayo2020,Rice2023}, these parameters have generally been treated as secondary diagnostics rather than incorporated systematically as primary features in machine-based classifications. Their potential role in shaping data-driven clustering outcomes within a fully probabilistic, high-dimensional framework therefore remains largely unexplored.

In this work, dynamically motivated parameters are incorporated into a two-stage Gaussian mixture model (GMM) to identify statistically supported clusters within the observed population of close-in giant-dominated exoplanets. The clustering is performed in a feature space dominated by derived dynamical quantities, including the Hill radius, the Safronov number, and the escape-to-orbital velocity ratio. These quantities are computed from planetary and stellar parameters provided by the NASA Exoplanet Archive and encode characteristic dynamical scales and interaction strengths of the planet--star system, rather than relying solely on directly observed properties. The resulting clusters are subsequently mapped onto a pebble-accreted synthetic population drawn from planet-formation simulations that explicitly track formation timing relative to disk dispersal, gas-envelope accretion efficiency, and solid composition \citep{Johansen2019,JohansenLyra2026}, allowing the observed exoplanet sub-populations to be interpreted in the context of formation pathways predicted by pebble-accretion models.

This paper is organised as follows. Section~\ref{data_select} describes the observational sample selection and the construction of the input feature space. Section~\ref{method} outlines the methodology, including the two-stage GMM clustering, clustering diagnostics, and the probabilistic mapping to the synthetic population. The clustering results are presented in Section~\ref{result}. In Section~\ref{discuss}, the physical interpretation of the clusters is discussed in the context of pebble-accretion-based formation pathways. Section~\ref{conclusion} summarises the main conclusions. Detailed descriptions of the input parameter validation, dimensionality-reduction tests, synthetic population construction, migration prescriptions, and additional statistical diagnostics are provided in the Appendices.

\section{Data selection}\label{data_select}

\subsection{Z-score normalisation and outlier filtering}

Planetary parameters were retrieved from the NASA Exoplanet Archive, and the sample was restricted to transiting planets hosted by single-star systems. The dataset comprises 2908 planets orbiting single-star hosts. Only planets with directly measured physical parameters were retained; in particular, planetary masses inferred from empirical mass--radius relations were excluded, and the measured mass ($M_{\rm p}$) was adopted. The working sample was then defined by requiring valid values for all nine adopted parameters as listed in Table~\ref{tab:feature_parameters}, namely the orbital period ($P_{\mathrm{orb}}$), planetary radius ($R_\mathrm{p}$), planetary surface gravity ($g_\mathrm{p}$), $\Theta_\mathrm{Saf}$, $R_\mathrm{Hill}$, planet--star mass ratio ($q_\mathrm{p/s}$), $v_\mathrm{esc}/v_\mathrm{orb}$, scaled semi-major axis ($a/R_\mathrm{s}$), and the number of confirmed planets in the system ($N_\mathrm{p}$), yielding 551 planets.

\begin{table}
\caption{Planetary and system parameters adopted in the GMM feature space.}
\label{tab:feature_parameters}
\centering
\setlength{\tabcolsep}{6pt}
\renewcommand{\arraystretch}{1.25}
\begin{tabular}{llp{5.7cm}}
\hline
Symbol & Unit & Description \\
\hline
$P_{\mathrm{orb}}$ 
& days
& Orbital period of the planet. \\

$R_{\mathrm{p}}$ 
& $R_{\mathrm{J}}$ 
& Planetary radius in Jupiter radius. \\

$g_{\mathrm{p}}$ 
& cm\,s$^{-2}$ 
& Planetary surface gravity. \\

$q_{\mathrm{p/s}}$ 
& -- 
& Planet--star mass ratio.\\

$R_{\mathrm{Hill}}$ 
& AU 
& Orbital Hill radius of the planet, defining the characteristic gravitational sphere of influence. \\

$\Theta_{\mathrm{Saf}}$ 
& -- 
& Safronov number, characterising the efficiency of gravitational scattering relative to orbital motion. \\

$v_{\mathrm{esc}}/v_{\mathrm{orb}}$ 
& -- 
& Ratio of planetary escape velocity to orbital velocity, tracing the balance between gravitational binding and orbital motion. \\

$a/R_{\mathrm{s}}$ 
& -- 
& Scaled semi-major axis, defined as the orbital semi-major axis in units of stellar radius. \\

$N_{\mathrm{p}}$ 
& -- 
& Number of confirmed planets in the host planetary system. \\
\hline
\end{tabular}
\end{table}

\begin{figure}
    \centering
    \includegraphics[width=0.49\textwidth]{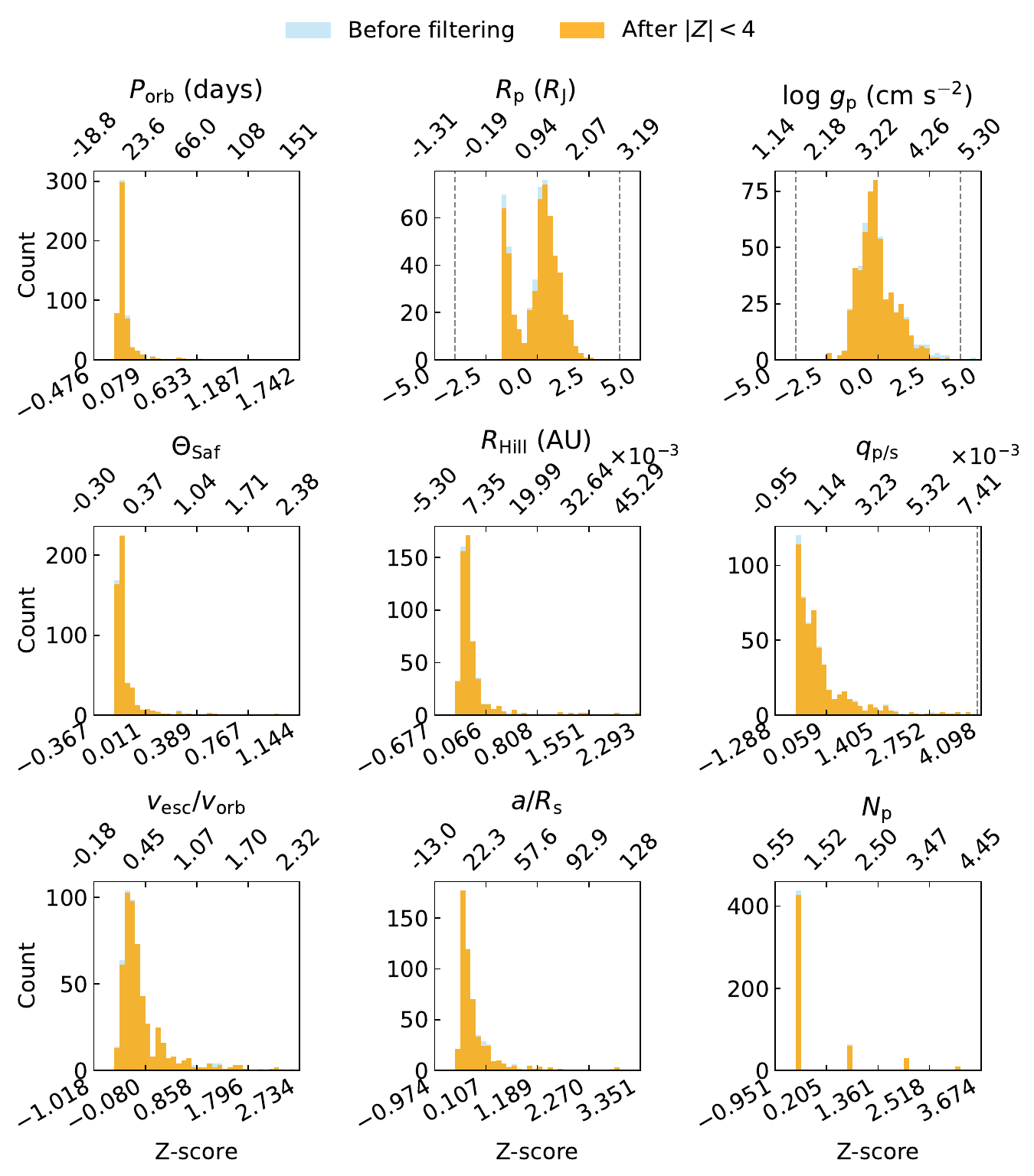}
    \caption{Distributions of the nine input parameters before (blue) and after (orange) Z-score filtering, shown in separate panels for each parameter. The bottom axis in each panel indicates the standardised value used in the GMM analysis, while the top axis shows the corresponding physical value in the original units (see Table~\ref{tab:feature_parameters}). Dashed vertical lines mark the adopted filtering threshold of $|Z| = 4$. The filtering removes a small number of extreme outliers while preserving the overall shape of the parameter distributions.
    }
    \label{fig1}
\end{figure}

All continuous parameters were standardised using Z-score normalisation, such that each parameter has zero mean and unit variance (Figure~\ref{fig1}). To reduce the influence of extreme outliers while preserving the overall shape of the distributions, a uniform filtering criterion of $|Z| < 4$ was applied, resulting in a final sample of 529 planets. Figure~\ref{fig1} shows the distributions of each parameter before and after filtering, demonstrating that this procedure removes only a small number of extreme values without significantly modifying the bulk of the parameter space.

\subsection{Planetary classification labels}

For interpretative purposes, we assign each planet a set of descriptive category labels based on simple cuts in $P_{\rm orb}$, $M_{\rm p}$, $R_{\rm p}$, and $T_{\rm eq}$. These categories broadly follow common conventions in the exoplanet literature and are not intended to represent sharp physical boundaries. Importantly, the labels are not used as input features in the machine-learning analysis; instead, they serve solely to aid the interpretation of cluster compositions by quantifying and visualising the relative contribution of different planet populations. The clustering itself is performed entirely in a continuous parameter space and remains independent of any pre-defined classifications. A complete list of the adopted categories and their defining criteria is provided in Table~\ref{tab:planet_categories}. The abbreviations correspond to ultra-hot Jupiters (UHJ), fast hot Jupiters (FHJ), hot Jupiters (HJ), warm Jupiters (WJ), cold Jupiters (CJ), hot Saturns (HS), hot Neptunes (HN), warm Neptunes (WN), and hot sub-Neptunes (HsN).

\begin{table}
\caption{Planet categories used for interpretative reference.}
\label{tab:planet_categories}
\centering
\setlength{\tabcolsep}{4.5pt}
\renewcommand{\arraystretch}{1.2}

\begin{threeparttable}
\begin{tabular}{lccccc}
\hline
Category & $P_{\rm orb}$ (d) & $R_\mathrm{p}$ ($R_\mathrm{J}$) & $M_\mathrm{p}$ ($M_\mathrm{J}$) & $T_\mathrm{eq}$ (K) & $N$ \\
\hline
UHJ   & 0--10    & 0.74--1.224  & <13 & $\ge$2000 & 28  \\
FHJ   & 0--3     & 0.74--1.224  & <13 & <2000    & 82  \\
HJ    & 3--10    & 0.74--1.224  & <13 & <2000    & 206 \\
WJ    & 10--100  & 0.74--1.224  & <13 & --       & 57  \\
CJ    & $\ge$100 & 0.74--1.224  & <13 & --       & 6   \\
HS    & 0--10    & 0.455--0.74  & --  & --       & 26  \\
HN    & 0--10    & 0.276--0.455 & --  & --       & 27  \\
WN    & 10--100  & 0.276--0.455 & --  & --       & 29  \\
HsN   & 0--10    & 0.17--0.276  & --  & --       & 68  \\
\hline
\end{tabular}

\begin{tablenotes}
\footnotesize
\item Notes: These labels are applied after the clustering and are not used as input features in the machine-learning analysis. Abbreviations: UHJ = ultra-hot Jupiter, FHJ = fast hot Jupiter, HJ = hot Jupiter, WJ = warm Jupiter, CJ = cold Jupiter, HS = hot Saturn, HN = hot Neptune, WN = warm Neptune, and HsN = hot sub-Neptune.
\end{tablenotes}

\end{threeparttable}
\end{table}

\subsection{Synthetic populations}

The synthetic populations adopted in this work are drawn from the planet formation simulations presented by \citet{Johansen2019} and \citet{JohansenLyra2026}, which are based on the pebble accretion framework coupled with disk evolution and planet migration. In this model, planetary cores grow through the accretion of mm–cm sized pebbles with Stokes numbers ${\rm St}=0.01$ that are embedded in a protoplanetary disk. Planet formation proceeds through the coupled processes of pebble accretion, gas-envelope growth, and orbital migration, with their relative efficiencies regulated by planetary mass and the evolving properties of the protoplanetary disk.

The simulations self-consistently track the coupled evolution of planetary mass, orbital distance, and disk gas content over time, and include prescriptions for gas accretion, gap opening, and Type~I/II migration. As a result, the synthetic planets are associated with a set of physically meaningful quantities that are not directly accessible from current observations. These include the fraction of disk gas remaining at the onset of planet formation, parameterised through the formation time ($tt0$), the partitioning of planetary mass between solid and gaseous components, quantified by the core mass ($M_{\mathrm{pla}}$) and gas-envelope mass ($M_{\mathrm{gas}}$), as well as composition-related metrics such as the ice mass ($M_{\mathrm{ice}}$) and the resulting ice--rock mass ratio. In addition, each synthetic system is characterised by its $g_{\mathrm{p}}$, $R_{\mathrm{Hill}}$, $q_{\mathrm{p/s}}$, and $a$, which provide complementary measures of dynamical scale and interaction strength. These synthetic populations therefore provide a physically motivated reference for interpreting the observed clustering results, enabling the observational sample to be placed in the context of underlying formation pathways rather than being interpreted solely in terms of present-day observables.

\section{Methodology}\label{method}

\begin{figure}
    \centering
    \includegraphics[width=0.49\textwidth]{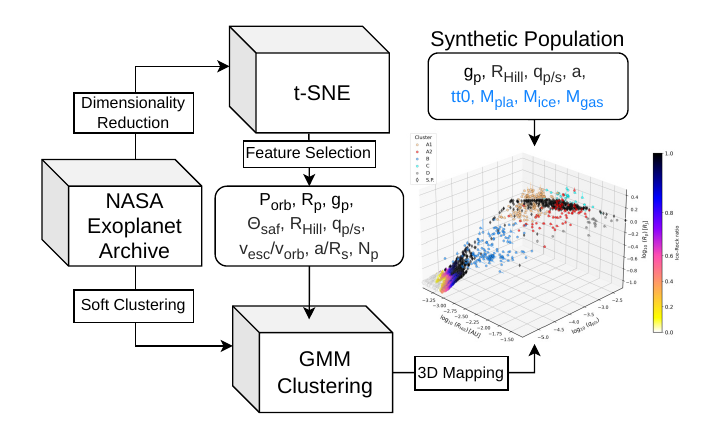}
    \caption{Schematic overview of the analysis pipeline adopted in this work, in which planets from the NASA Exoplanet Archive are clustered using machine-learning techniques and subsequently mapped into a three-dimensional parameter space using a simulation-based synthetic population. Parameters shown in blue denote quantities that are not directly accessible from current observations and are therefore defined only within the synthetic population, including the planetary surface gravity ($g_{\mathrm{p}}$), the Hill radius ($R_{\mathrm{Hill}}$), the planet–star mass ratio ($q_{\mathrm{p/s}}$), the semi-major axis ($a$), the planet formation time ($tt0$), the core mass ($M_{\mathrm{pla}}$), the ice mass ($M_{\mathrm{ice}}$), and the gas-envelope mass ($M_{\mathrm{gas}}$).}\label{fig2}
\end{figure}

The methodology combines unsupervised machine-learning techniques with physically motivated parameters to investigate the population-level organisation of the observed exoplanet population and to relate it to formation pathways inferred from synthetic populations. Figure~\ref{fig2} presents a schematic overview of the analysis pipeline adopted in this study.

This study adopts an unsupervised learning approach to identify natural groupings in the observed exoplanet population based solely on measured physical and orbital properties, without relying on predefined labels or training examples. The analysis is based on the planet sample described in Section~\ref{data_select}, characterised by a selected set of parameters that define the input feature space for clustering. In addition to directly reported observables, the feature space includes several derived dynamical quantities, namely the orbital Hill radius ($R_{\mathrm{Hill}}$), the scaled semi-major axis ($a/R_\mathrm{s}$), the planetary surface gravity ($g_\mathrm{p}$), the Safronov number ($\Theta_{\mathrm{Saf}}$), and the escape-to-orbital velocity ratio ($v_{\mathrm{esc}}/v_{\mathrm{orb}}$). These quantities are computed from the planetary and stellar parameters provided by the NASA Exoplanet Archive and encode characteristic dynamical scales and scattering efficiencies of the planet--star system; their definitions and derivations are given in Appendix~\ref{app:dynamical_quantities}.

As an initial diagnostic step, t-distributed stochastic neighbour embedding (t-SNE) is applied to visualise the distribution and relationships within the data and to assess the presence of potential groupings~\citep{Maaten2008}. A detailed evaluation of the reliability and stability of the t-SNE embedding is provided in Appendix~\ref{tSNE_append}. The t-SNE projection is used exclusively for exploratory visualisation and feature-selection diagnostics and does not enter the clustering analysis. The final set of input parameters adopted for the clustering is motivated and validated through a suite of complementary diagnostic tests described in Appendix~\ref{input_append}, which demonstrate that the selected features provide a physically meaningful and internally coherent representation of the planetary parameter space.

The primary classification is performed using a Gaussian mixture model (GMM), which enables probabilistic clustering in the original multi-dimensional feature space. This framework is well suited to exoplanet populations, whose formation and evolutionary pathways are expected to form continuous distributions rather than discrete classes~\citep{Jiang2006}, allowing individual planets to occupy intermediate states between clusters without imposing hard boundaries.

The quality and robustness of the resulting partitions are evaluated using standard clustering diagnostics commonly adopted in astronomical machine-learning studies~\citep{Feinstein2024,Semenov2025}, including the silhouette (SIL) score, the Davies--Bouldin index (DBI), and the Calinski--Harabasz index (CHI). The SIL score quantifies the balance between within-cluster cohesion and separation from neighbouring clusters, with larger values indicating better-defined groupings. The DBI measures the ratio of intra-cluster scatter to inter-cluster separation and is minimised for compact, well-separated clusters; values of order unity or lower are commonly taken to indicate moderate to good cluster separation. The CHI evaluates the relative dispersion between and within clusters and is typically maximised when clusters are compact and well separated. Together, these diagnostics provide complementary measures of cluster separation, compactness, and statistical support.

While the GMM is trained in a Z-score–standardised feature space to ensure numerical stability and balanced weighting of the input parameters, the subsequent KDE-based mapping is performed in a logarithmic physical parameter space to facilitate the interpretation of the resulting populations across parameters that span multiple orders of magnitude. Specifically, the observed planets and the synthetic population are jointly embedded in a three-dimensional space defined by $R_\mathrm{p}$, $q_\mathrm{p/s}$, and $R_\mathrm{Hill}$, which together provide a representative basis for the dominant variance captured by the leading principal components (see Appendix~\ref{input_append}). The construction of the synthetic populations and the adopted prescription for late-stage migration are described in Appendix~\ref{SP_append}. Embedding both the observational data and the synthetic systems in the same three-dimensional physical space enables a direct and internally consistent comparison, allowing the data-driven clustering results to be interpreted in terms of underlying formation conditions.

To facilitate this interpretation, the synthetic population is characterised using a set of physically motivated parameters that encode complementary aspects of planet formation. In this work, we focus on three such quantities: the gas availability parameter $\mathcal{G}_1$, which traces formation timing relative to disk dispersal; the planetary gas mass fraction $f_{\mathrm{gas}}$, which traces envelope accretion efficiency; and the ice--rock mass ratio, which encodes the composition of the solid component and hence the formation environment within the protoplanetary disk.

The parameter $\mathcal{G}_1$ quantifies the fraction of disk gas remaining at the onset of planet formation. It is defined assuming an exponential decay of the disk gas mass with time,
\begin{equation}
\mathcal{G}_1(tt0) = \exp\!\left(-\frac{tt0}{\tau_{\mathrm{disk}}}\right),
\end{equation}
where $tt0$ denotes the time at which planet formation begins and $\tau_{\mathrm{disk}} \simeq 2.5\,\mathrm{Myr}$ is the characteristic disk dispersal timescale~\citep{Mamajek2009}. Larger values of $\mathcal{G}_1$ correspond to earlier formation epochs in more gas-rich disks, while smaller values indicate formation at later stages when the gaseous component has been substantially depleted.

The gas mass fraction $f_{\mathrm{gas}}$ is defined as the ratio of gaseous envelope mass to the total planetary mass. The ice--rock mass ratio describes the relative contribution of volatile-rich and refractory solids in the planetary building material. Together, these parameters provide a physically interpretable basis for probabilistically associating synthetic planets with the GMM-defined clusters of observed planets, thereby linking the data-driven clustering to formation timing, envelope growth, and solid composition in a unified framework.

\section{Results}\label{result}

\subsection{Two-stage Gaussian mixture model clustering}\label{GMM_two_stage_results}

\begin{figure}
    \centering
    \includegraphics[width=0.49\textwidth]{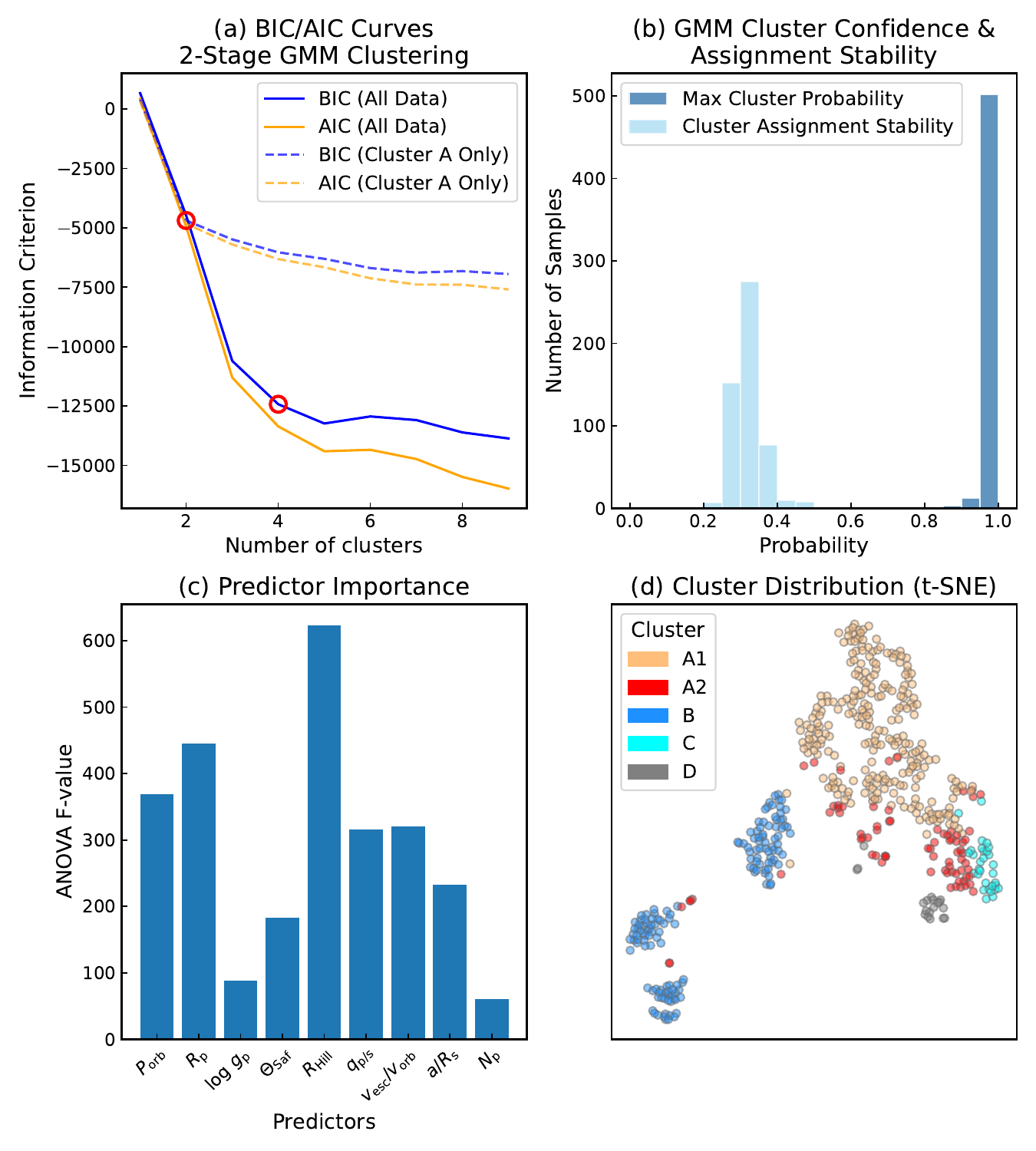}
    \caption{Statistical diagnostics and characterisation of the two-stage GMM clustering. (a) Bayesian Information Criterion (BIC) and Akaike Information Criterion (AIC) as a function of the number of mixture components for the first-stage GMM applied to the full sample (solid lines), and for the second-stage GMM applied exclusively to cluster~A (dashed lines). Open red circles mark the locations where a clear change in slope is observed in the information criteria, corresponding to the preferred number of mixture components adopted at each stage. The second-stage clustering is performed to further resolve the dominant population within cluster~A. (b) Distribution of the maximum posterior cluster membership probability for each object (dark blue), together with the cluster assignment stability estimated from bootstrap resampling (light blue), defined as the fraction of resampled realisations in which each object retains the same cluster label. (c) Predictor importance for the final five-cluster solution, quantified by ANOVA F-values, illustrating the relative contribution of each orbital and physical parameter to the cluster separation. (d) Two-dimensional t-SNE projection of the final clustering outcome, used solely for visualisation of the high-dimensional feature space. The axes correspond to the two embedding dimensions produced by the t-SNE algorithm and do not have direct physical meaning. Colours indicate the final GMM cluster labels (A1, A2, B, C, and D) and are consistent across panels.
    }\label{fig3}
\end{figure}

A two-stage GMM clustering strategy was applied to the planetary sample, with the statistical diagnostics and clustering outcomes summarised in Figure~\ref{fig3}. The optimal number of mixture components was assessed using the Bayesian Information Criterion (BIC) and the Akaike Information Criterion (AIC). In the first stage, both criteria favour a four-component solution, with the solid curves in Figure~\ref{fig3}a exhibiting a clear change in slope around four, motivating the adoption of a four-cluster partition (A--D).

The first-stage GMM yields a statistically supported partition in the original feature space, with clustering diagnostics (SIL = 0.376, DBI = 0.929, CHI = 202.8) indicative of a moderately separated, non-random clustering with partial overlap between components. Cluster~A contains the majority of the sample (62.9\%), motivating a second-stage GMM applied exclusively to this subset.

Based on the dashed BIC and AIC curves in Figure~\ref{fig3}a, which show a clear change in slope at two, the second-stage refinement subdivides cluster~A into two components, A1 and A2. Following this subdivision, the global clustering metrics change to SIL = 0.078, DBI = 1.970, and CHI = 183.5, reflecting the increased overlap expected when resolving finer internal distinctions within a previously dominant cluster. Despite the reduced global separation in purely geometric terms, the subdivision is supported by systematic differences in dynamical parameters, as indicated by the ANOVA F-values. The final clustering thus consists of five components (A1, A2, B, C, and D).

The robustness of the GMM classifications is illustrated in Figure~\ref{fig3}b. The maximum posterior cluster membership probability exceeds 0.8 for most objects, indicating that individual planets are typically well associated with a dominant mixture component in a given realisation of the GMM. In contrast, the cluster assignment stability estimated via bootstrap resampling is typically distributed in the range 0.2–0.4, reflecting that a subset of objects may switch between neighbouring clusters across resampled datasets. This behaviour is expected for a probabilistic clustering applied to partially overlapping populations. Figure~\ref{fig3}c shows the relative importance of each input parameter for the final five-cluster solution, quantified using Analysis of Variance (ANOVA) F-values. The largest contributions arise from $R_{\mathrm{Hill}}$ and $R_\mathrm{p}$, whereas $N_\mathrm{p}$ and $\log g_\mathrm{p}$ contribute more weakly to the cluster separation. The final clustering outcome is visualised in Figure~\ref{fig3}d using a 2D t-SNE projection, with points coloured according to their GMM cluster assignments. This projection is used solely for visualisation, while the clustering itself is performed in the full multi-dimensional feature space.

\subsection{Cluster characterisation and planet category mapping}\label{sec:cluster_char}

\begin{figure}
    \centering
    \includegraphics[width=0.49\textwidth]{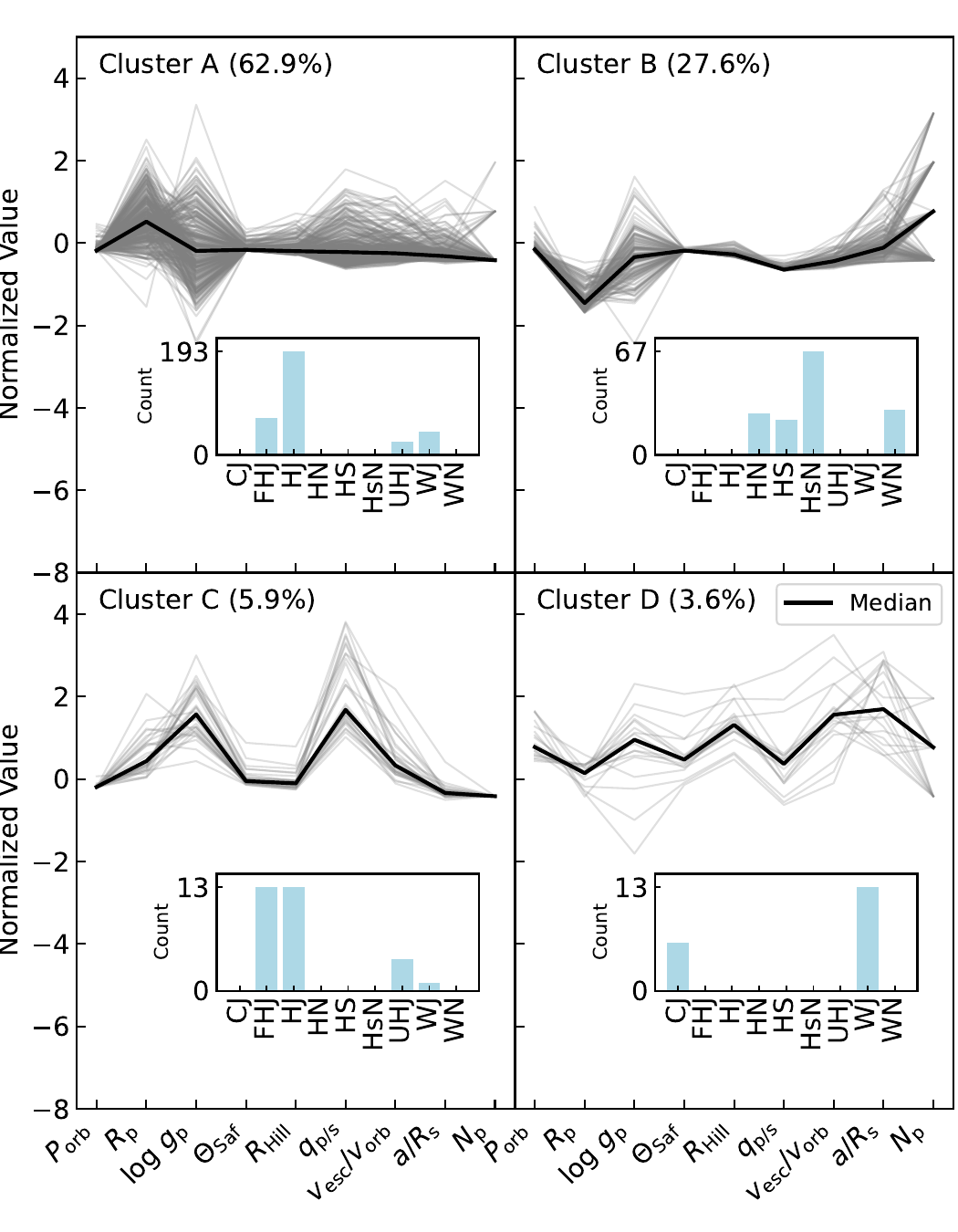}
    \caption{Normalised parameter profiles for the four clusters (A–D) identified in the first-stage GMM classification. In each panel, grey curves show the normalised parameter vectors of individual planets assigned to the corresponding cluster, while the black solid line indicates the median profile of the cluster. The percentage quoted in each panel title denotes the fraction of the full sample assigned to that cluster. Insets show the distribution of adopted planet categories within each cluster, illustrating how the GMM-defined regions of parameter space map onto commonly used planet classes.
    }\label{fig4}
\end{figure}

The properties of the GMM clusters are characterised using their normalised parameter profiles and their distribution across commonly used planet categories. Figure~\ref{fig4} presents the results of the first-stage GMM classification, showing the normalised profiles for the four primary clusters (A--D). In each panel, the median profile and the dispersion among individual planets illustrate differences in orbital and physical parameters between clusters, together with the fraction of the full sample assigned to each group.

\begin{figure}
    \centering
    \includegraphics[width=0.49\textwidth]{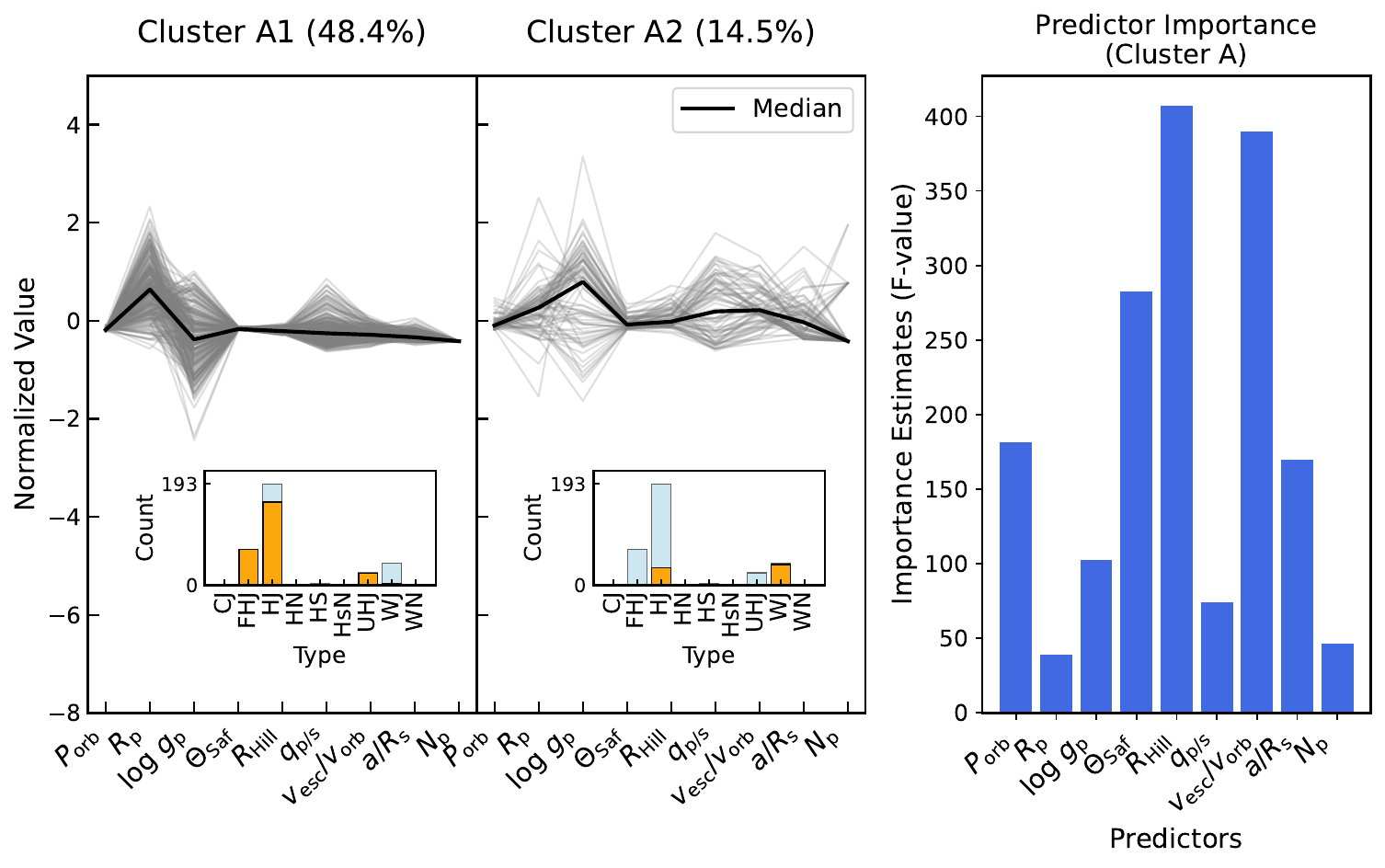}
    \caption{Normalised parameter profiles and predictor importance for the second-stage GMM subdivision of cluster~A into sub-clusters~A1 and~A2. Left and middle panels show the normalised parameter profiles for clusters~A1 and~A2, respectively. Grey curves represent individual planets, while the black solid line indicates the median profile of each sub-cluster. The percentage quoted in each panel title denotes the fraction of the full sample assigned to that sub-cluster. Insets display the distribution of descriptive planet categories: light-blue bars correspond to the category distribution of the parent cluster~A, while orange bars show the distributions within sub-clusters~A1 and~A2, illustrating how the second-stage GMM further partitions the dominant cluster. The right-hand panel shows the predictor importance for the subdivision of cluster~A, quantified by ANOVA F-values, highlighting the parameters that most strongly contribute to the separation between A1 and A2 in the reduced parameter space.
    }\label{fig5}
\end{figure}

Figure~\ref{fig5} focuses on the second-stage subdivision of cluster~A into sub-clusters~A1 and~A2. The normalised profiles of these sub-clusters reveal differences within the originally dominant population, while the associated category distributions show how the second-stage GMM further partitions planets that were grouped together in the first-stage classification. A comparison between Figures~\ref{fig4} and~\ref{fig5} highlights the distinct roles of the two clustering stages: the first-stage GMM separates the full sample into clusters A--D with clearly distinct parameter profiles and category compositions, whereas the second-stage subdivision of cluster~A yields more modest differences consistent with refinement within a single population.

\begin{figure}
    \centering
    \includegraphics[width=0.49\textwidth]{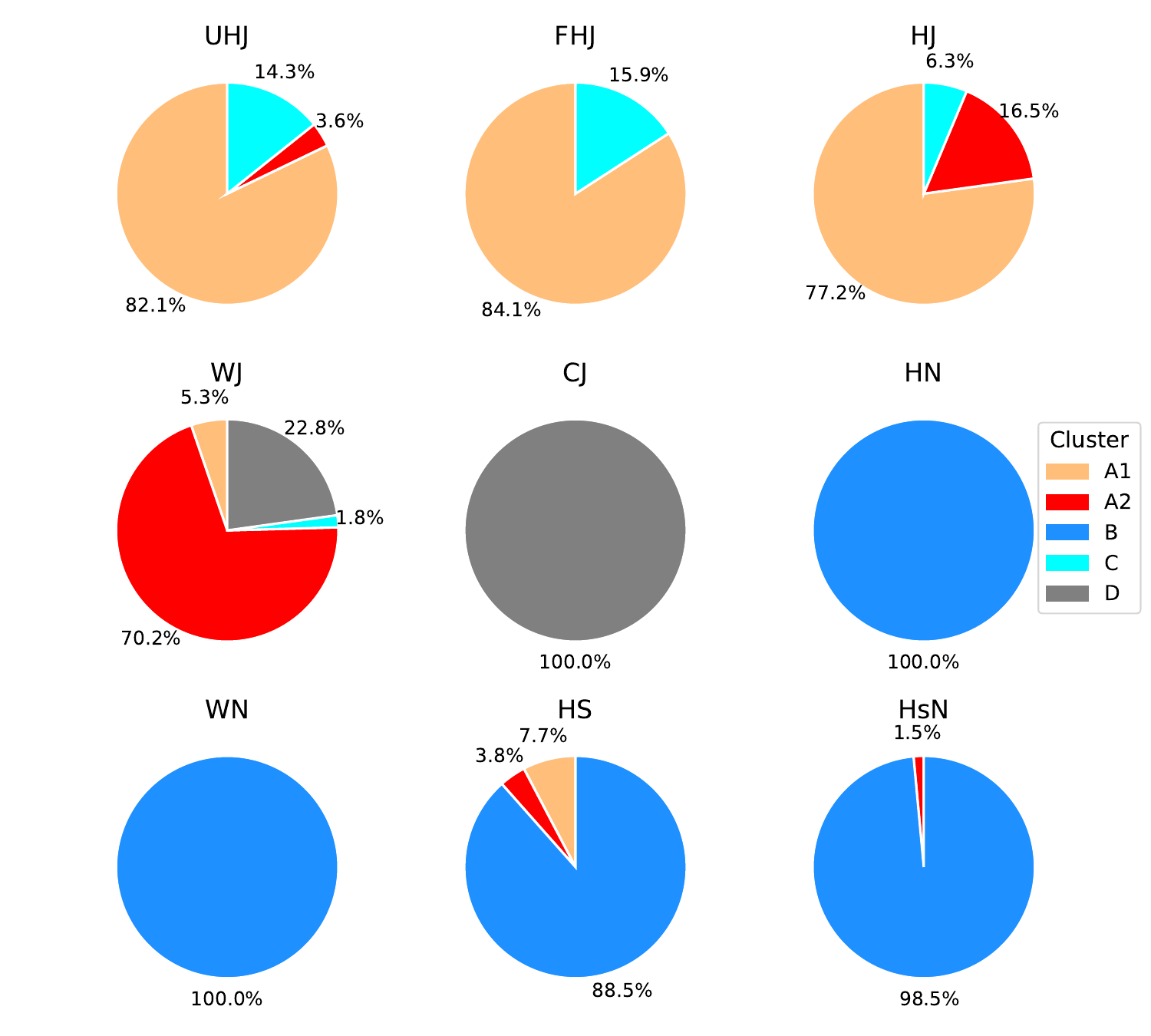}
    \caption{Relative contributions of the final GMM clusters (A1, A2, B, C, and D) within Table~\ref{tab:planet_categories} defined planet category. Each pie chart shows the fractional cluster composition for a given category. Percentages indicate the fraction of objects in each category assigned to a given cluster.
    }\label{fig6}
\end{figure}

Despite the relatively subtle separation in parameter space, the second-stage clustering produces a redistribution of planet categories. In particular, WJs and a subset of HJs are preferentially assigned to sub-cluster~A2, which comprises 14.5\% of the full sample. The associated ANOVA F-values show that the subdivision is driven primarily by dynamical parameters, notably $R_{\mathrm{Hill}}$, $v_{\mathrm{esc}}/v_{\mathrm{orb}}$, and $\Theta_{\mathrm{Saf}}$, while $R_{\mathrm{p}}$ and $N_{\mathrm{p}}$ contribute comparatively little to the A1--A2 distinction. To facilitate interpretation, the cluster labels are retained in their neutral form (A1, A2, B, C, and D), as the classification is determined entirely by the unsupervised GMM. Figure~\ref{fig6} illustrates how these machine-learning-defined clusters map onto commonly used descriptive planet categories in a post hoc manner.

Several clear trends emerge from this comparison. UHJs, FHJs, and HJs are dominated by cluster~A1. Sub-cluster~A2 constitutes the dominant population among warm Jupiters and represents the second-largest contribution within the HJ category. CJs are exclusively assigned to cluster~D, which also contains a small fraction of WJs. Lower-mass planet classes, including HN, WN, HS, and HSubN, are predominantly associated with cluster~B. The HS category additionally includes a minor contribution from clusters~A1 and~A2. Figure~\ref{fig6} demonstrates that recognisable planet categories emerge from the unsupervised clustering despite the absence of categorical information in the clustering inputs.

\begin{figure}
    \centering
    \includegraphics[width=0.49\textwidth]{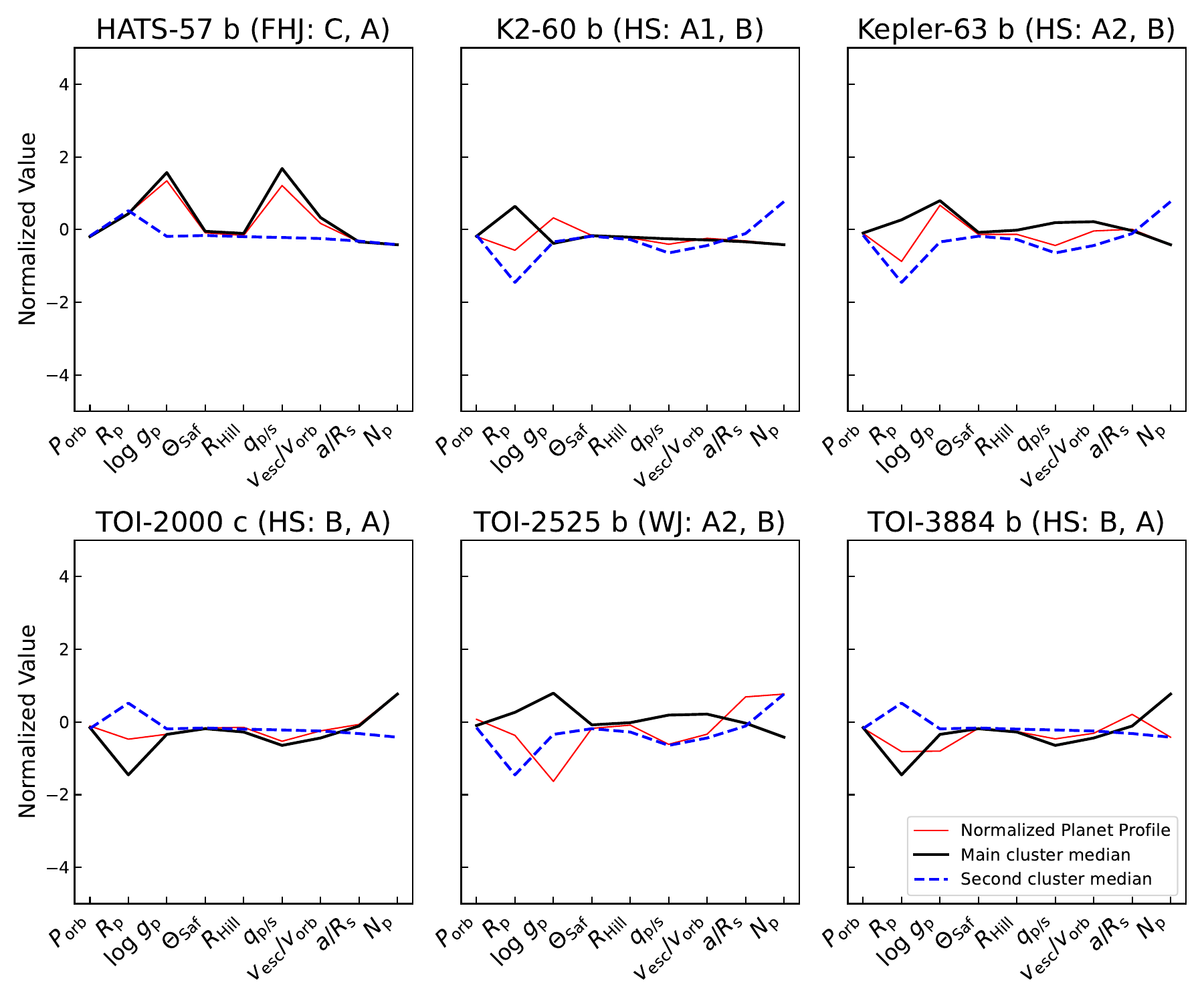}
    \caption{Normalised parameter profiles for six planets classified as low-confidence cases in the GMM analysis. For each object, the red curve shows the normalised planetary parameter vector, while the black solid line indicates the median profile of the primary GMM cluster. The blue dashed line corresponds to the median profile of the second most probable cluster based on posterior membership probabilities.
    }\label{fig7}
\end{figure}

A small number of objects exhibit ambiguous cluster membership. Figure~\ref{fig7} highlights these low-confidence cases, defined as planets with a maximum posterior cluster membership probability below 0.7. For these objects, comparable probabilities are assigned to more than one cluster, indicating proximity to the boundaries between GMM components in the multi-dimensional parameter space. In several cases, the first-stage GMM assigns these planets to a non-A cluster, although cluster~A appears as the second most probable association. As the second-stage refinement is restricted to cluster~A, secondary associations for low-confidence cases necessarily involve cluster~A and its sub-components; Figure~\ref{fig7} therefore compares each object with the median properties of both its primary and alternative assignments.

\subsection{Physical distributions and low-confidence cluster assignments}

\begin{figure}
    \centering
    \includegraphics[width=0.49\textwidth]{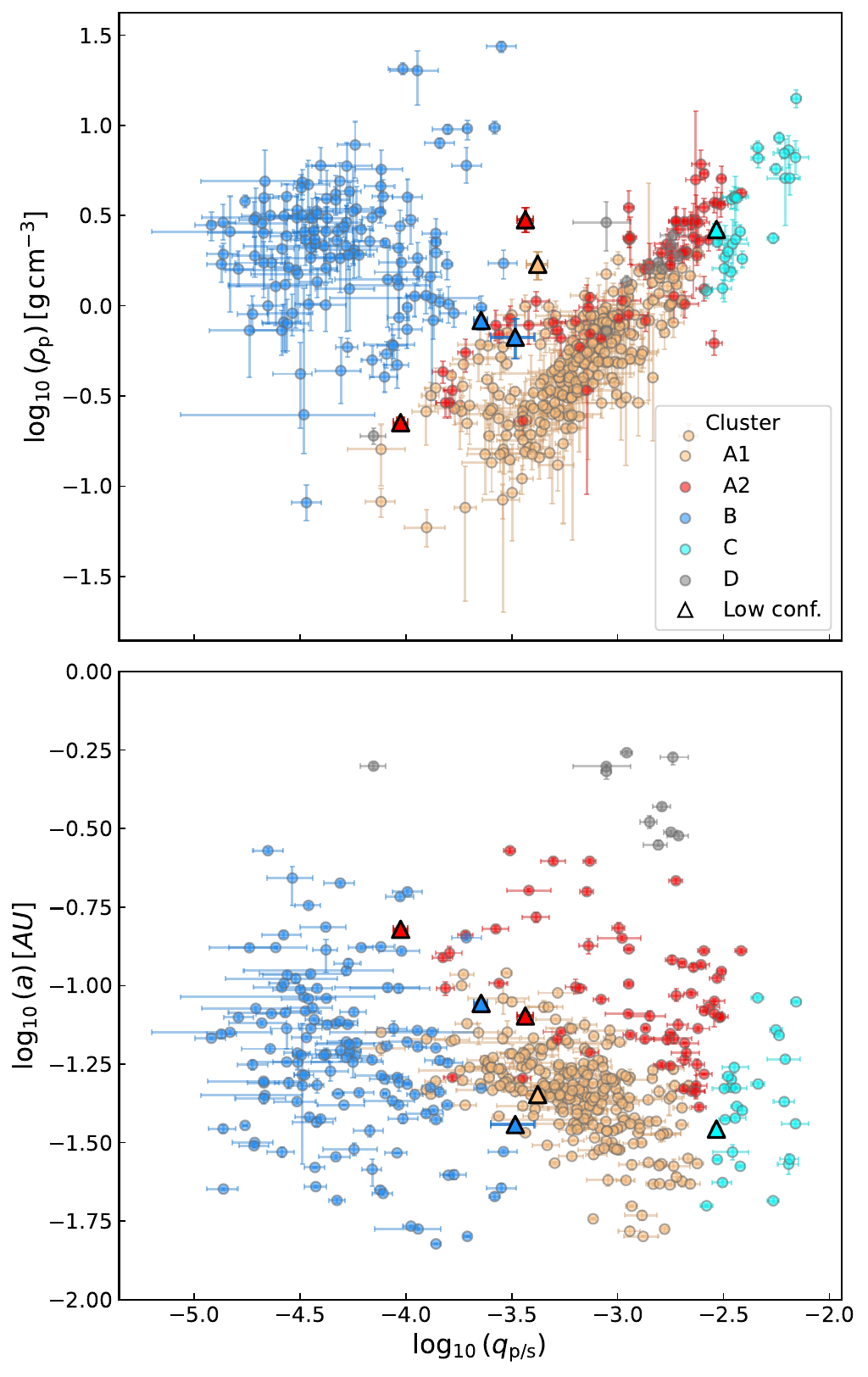}
    \caption{Logarithmic plane of $\rho_\mathrm{p}$--$q_\mathrm{p/s}$ (Top) and $a$--$q_\mathrm{p/s}$ (Bottom). Filled circles denote planets from the NASA Exoplanet Archive, coloured by their GMM cluster assignments (A1, A2, B, C, and D), with error bars indicating measurement uncertainties. Planets with low clustering confidence are marked by triangles.
    }\label{fig8}
\end{figure}

To further characterise the GMM clustering results, the cluster assignments are examined in several physical parameter spaces, including quantities that were not used as inputs to the clustering. Figures~\ref{fig8} shows the distribution of clusters in these physical planes, with planets exhibiting low clustering confidence explicitly highlighted.

In the logarithmic $\rho_\mathrm{p}$--$q_\mathrm{p/s}$ plane (Figure~\ref{fig8}, top), where $\rho_\mathrm{p}$ denotes the mean planetary density, cluster~C is primarily associated with large mass ratios, while cluster~B occupies a regime of lower masses and higher densities. Clusters~A1, A2, and~D show substantial overlap in this projection. In contrast, a clearer separation between these clusters is visible in the $a$--$q_\mathrm{p/s}$ plane (Figure~\ref{fig8} Bottom), where A1, A2, and~D occupy largely distinct orbital-distance regimes with limited overlap. Planets classified as low-confidence cases are preferentially located near the boundaries between these regions, particularly in the orbital-distance projection.

The physical relevance of the A1–A2 subdivision is further supported by their separation in parameter planes not used for clustering, as demonstrated in Appendix~\ref{val_gmm_2nd}.

\subsection{Mapping GMM clusters onto synthetic population formation space}

\begin{figure}
    \centering
    \includegraphics[width=0.49\textwidth]{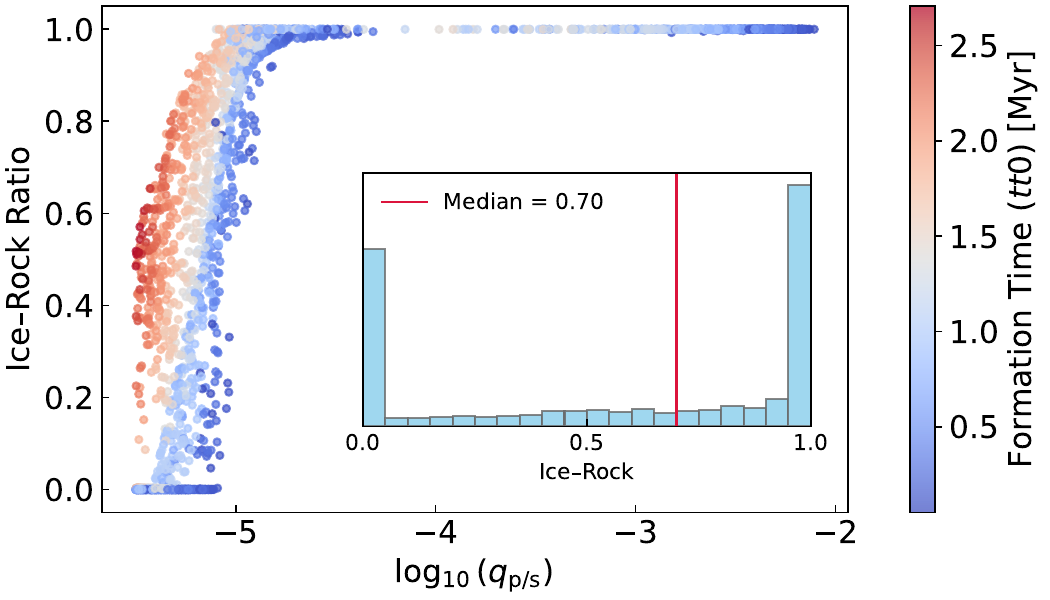}
    \caption{Distribution of synthetic planets in the $\log_{10}(q_{p/s})$–Ice–Rock ratio space. Each point represents a synthetic planet, color-coded by its formation time ${tt0}$ in Myr. The inset panel shows the normalised histogram of Ice–Rock ratios for the entire population, with the median value indicated by the red vertical line. The distribution is strongly bimodal, peaking near 0 and 1, corresponding to rock-dominated and ice-dominated compositions, respectively.
    }\label{fig9}
\end{figure}

A physically meaningful interpretation of the mapping between the observationally defined GMM clusters and the synthetic population first requires understanding how key formation-related properties are distributed within the synthetic population itself. Figure~\ref{fig9} therefore presents the distribution of synthetic planets in the $\log_{10}(q_{\mathrm{p/s}})$–ice–rock mass ratio plane, colour-coded by formation time $tt0$. This diagram illustrates the range of solid compositions and formation epochs produced by the population-synthesis model and provides a reference view of the formation parameter space spanned by the synthetic population.

\begin{figure}
    \centering
    \includegraphics[width=0.49\textwidth]{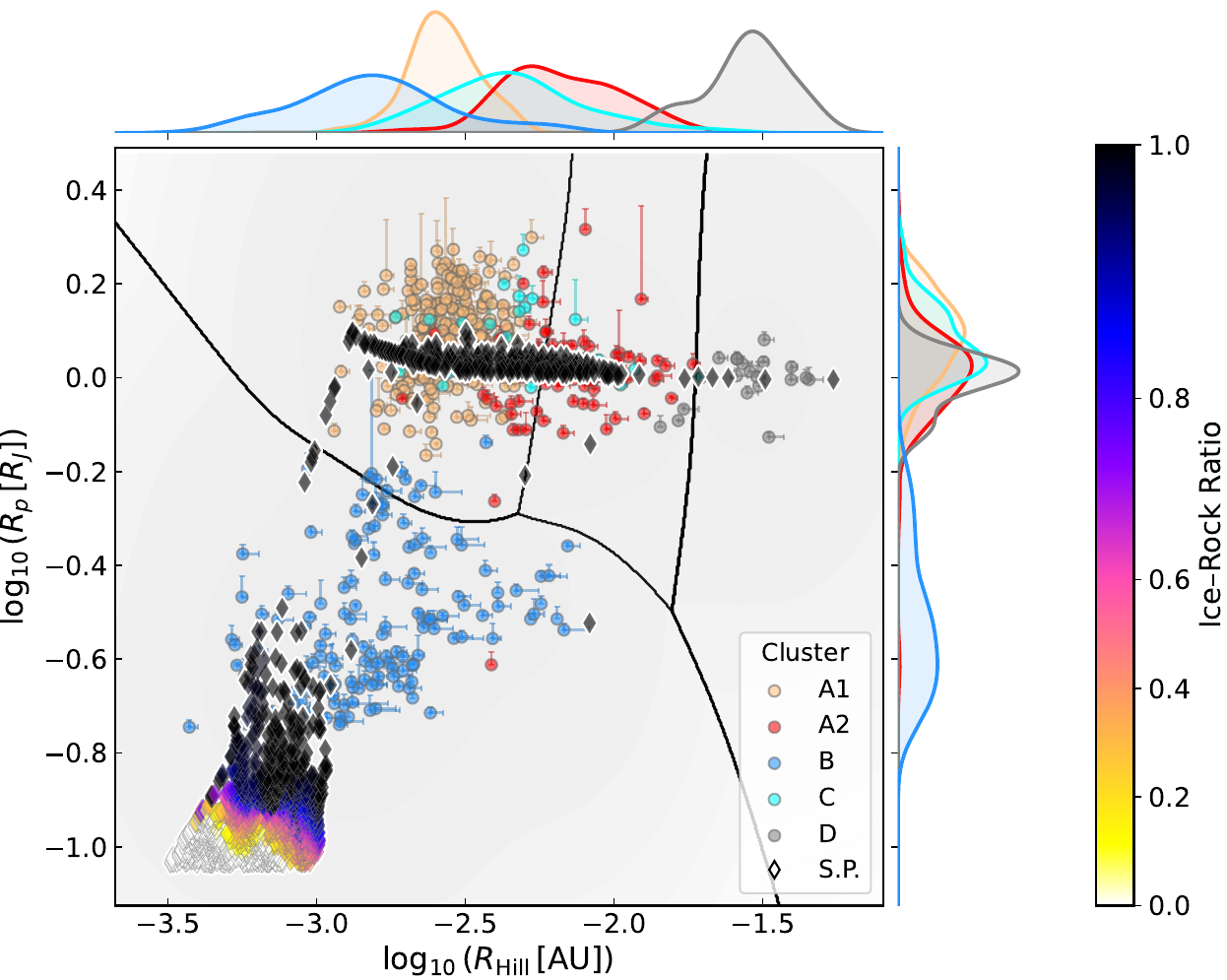}
    \caption{Mapping synthetic populations and GMM clusters onto the 2D logarithmic plane of $R_{\mathrm{Hill}}$--$R_\mathrm{p}$ , this figure shows the distribution of planets in terms of dynamical scale and planetary size.  Data are coloured according to the GMM clusters (A1, A2, B, C, and D), with error bars showing measurement uncertainties. Diamond symbols represent synthetic populations, coloured by their ice--rock mass ratio as indicated by the colour bar, which is scaled relative to the minimum, median, and maximum values of the synthetic populations. Black curves mark cluster decision boundaries derived from KDE in the  plane. Marginal kernel density distributions for each cluster are shown along the top and right axes.
    }\label{fig10}
\end{figure}

\begin{figure}
    \centering
    \includegraphics[width=0.49\textwidth]{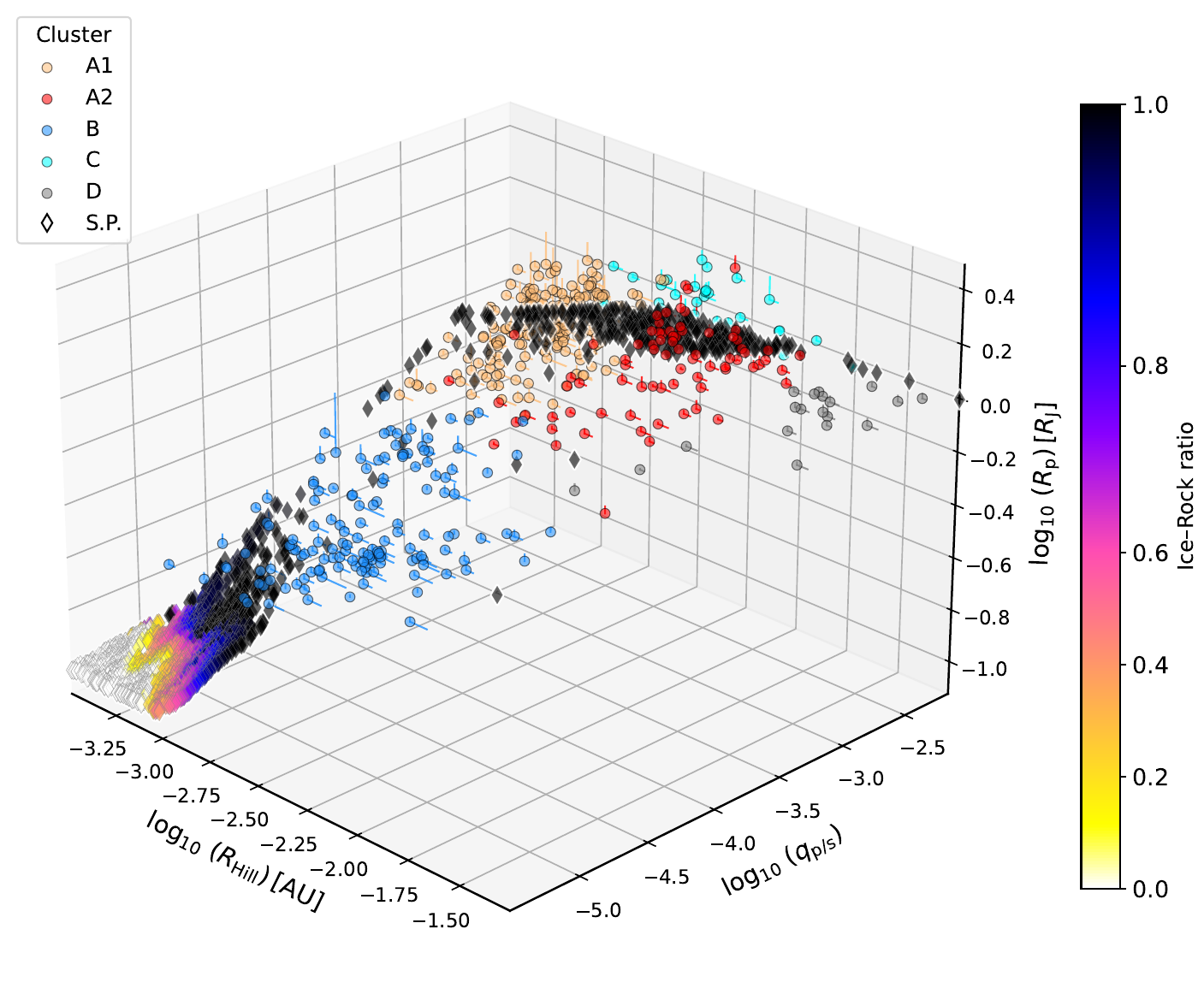}
    \caption{Distribution of observed planets and synthetic populations in the 3D logarithmic space defined by $R_{\mathrm{Hill}}$, $q_\mathrm{p/s}$, and $R_\mathrm{p}$. The colours and symbols follow the same definitions as in Figure~\ref{fig10}. The associated movie is available online.
    }\label{fig11}
\end{figure}

\begin{figure}
    \centering
    \includegraphics[width=0.49\textwidth]{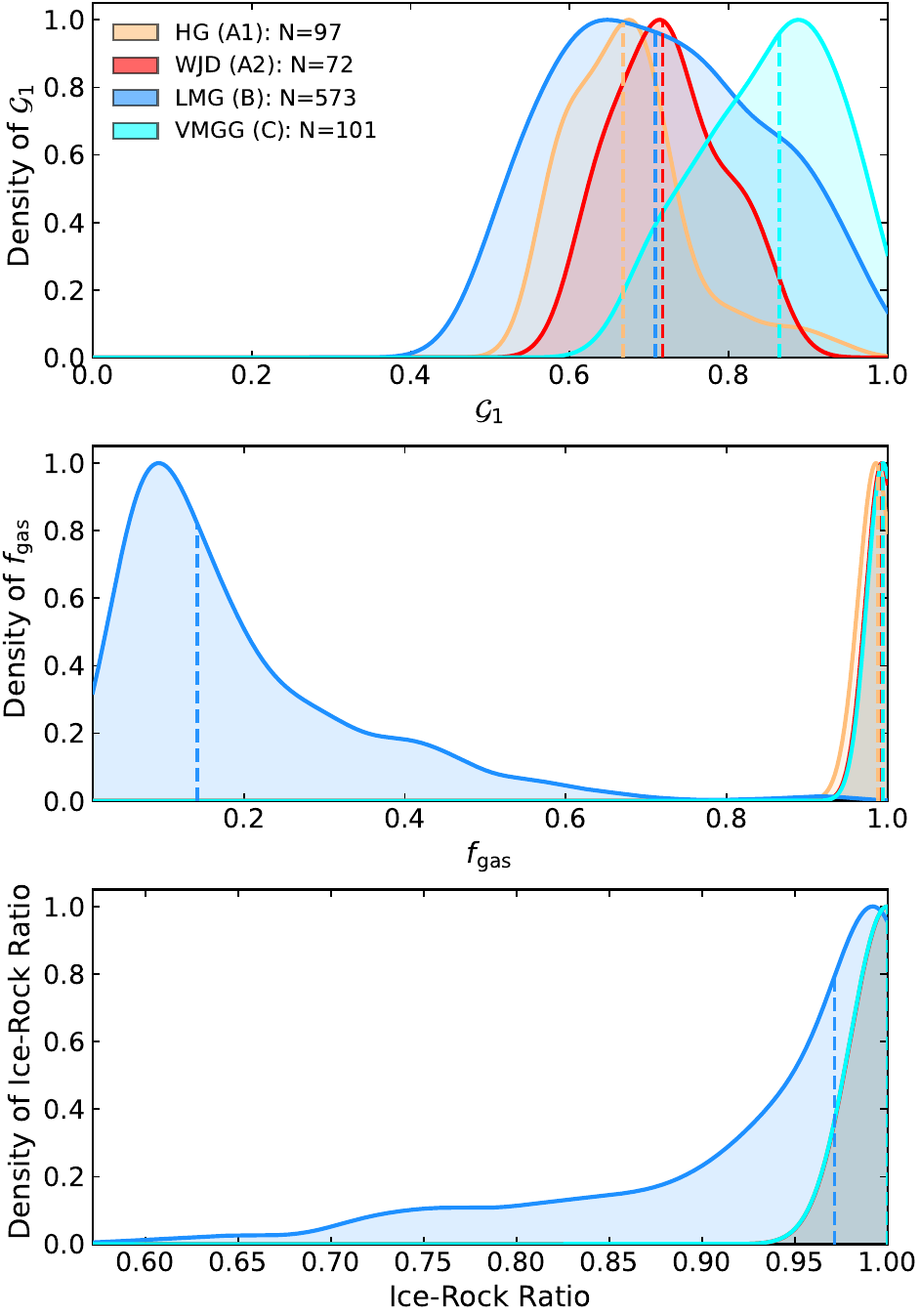}
    \caption{Kernel density estimates illustrating the distributions of three formation-related parameters for planets associated with the four clusters: HG (A1), WJD (A2), LMG (B), and VMGG (C). The parameters shown are the gas availability parameter $\mathcal{G}_1$, the gas mass fraction $f_{\mathrm{gas}}$, and the ice--rock mass ratio. For clarity, only synthetic planets within the overlap region of the observed parameter space (as defined in Appendix~\ref{app:overlap_robustness}) are included. The densities are normalised to a common peak value to facilitate comparison between clusters. Dashed vertical lines indicate the median value of each distribution.
    }
    \label{fig12}
\end{figure}

\begin{figure}
    \centering
    \includegraphics[width=0.49\textwidth]{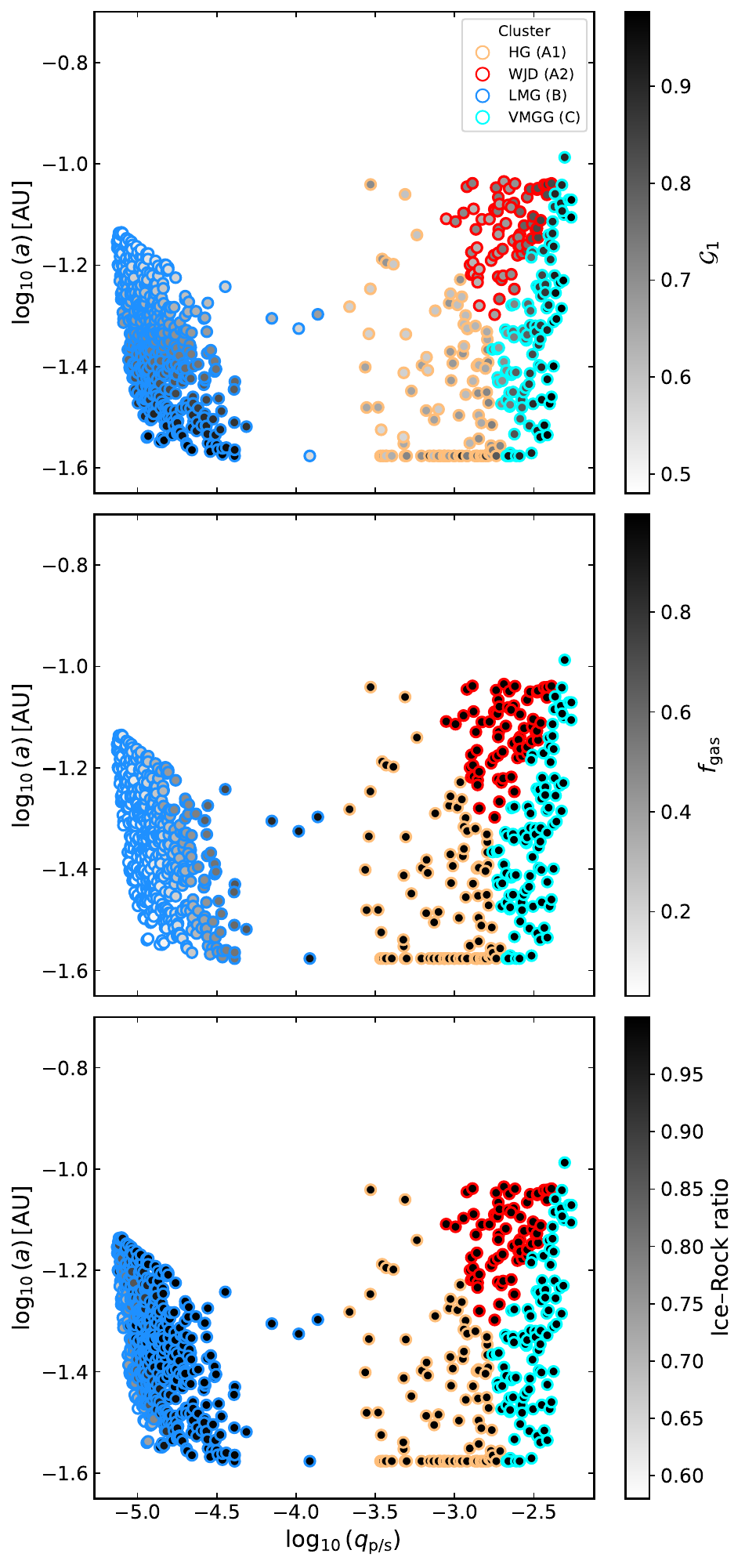}
    \caption{From top to bottom, the panels show synthetic populations mapped into the clustered parameter space in the logarithmic $a$--$q_\mathrm{p/s}$ planes, with inner colours indicating different formation-related properties: $\mathcal{G}1$ (Top), $f_\mathrm{gas}$ (Middle), and ice–rock mass ratio (Bottom). In all panels, edge colours indicate the GMM cluster assignments (A1, A2, B, and C) inferred from data, while the colour bars show the corresponding values of each parameter for the synthetic populations.
    }\label{fig13}
\end{figure}

Building on this synthetic baseline, the observationally defined GMM clusters are interpreted in the context of the same formation-related parameters available in the simulation. By associating observed cluster members with corresponding regions of the synthetic population in parameter space, it becomes possible to examine how the data-driven clusters relate to model-predicted formation conditions. Figures~\ref{fig10}–\ref{fig13} therefore compare the observed clusters and synthetic systems across parameter spaces defined by dynamical scale, planetary size, and selected formation quantities, enabling a direct connection between the GMM-based classification of observed planets and the formation properties predicted by the population-synthesis model.

Figure~\ref{fig10} presents the distribution of observed planets and synthetic populations in the logarithmic $R_{\mathrm{Hill}}$–$R_\mathrm{p}$ plane. These parameters are selected because they rank among the most discriminating features in the GMM classification, as indicated by their high ANOVA F-values (Figure~\ref{fig3}c). The Hill radius is defined following Equation~\ref{eq:mass_ratio}. Synthetic populations are overlaid as diamond symbols, with radii inferred from mass–radius relations (Appendix~\ref{app:transit_probability}). The colour coding of the synthetic populations represents their ice–rock mass ratio. Notably, the synthetic populations mapped onto cluster B preferentially exhibit lower volatile content, consistent with the lower-mass, higher-density nature of this cluster (Figure~\ref{fig8} top).

Because a 2D projection compresses higher-dimensional parameter space, Figure~\ref{fig11} extends this comparison to the 3D logarithmic space defined by $( R_{\mathrm{Hill}}, q_{\mathrm{p/s}},R_\mathrm{p})$. In this representation, cluster D is excluded from subsequent analysis owing to the limited number of observed systems and the negligible representation of synthetic counterparts.

The mapping between the observationally defined GMM clusters and the synthetic population is performed using a probabilistic approach based on kernel density estimation (KDE) in a three-dimensional logarithmic parameter space defined by $R_{\mathrm{Hill}}, q_{\mathrm{p/s}},$ and $R_{\mathrm{p}}$. For each observed cluster, a multivariate Gaussian KDE is constructed from the corresponding subset of the data, assuming uniform cluster priors. Synthetic planets are projected into the same parameter space and assigned posterior cluster associations based on the maximum KDE likelihood. To ensure that the inferred formation properties are not biased by synthetic systems located outside the observational domain, we additionally restrict the analysis to synthetic planets lying within the multivariate overlap region of each cluster, defined using a Mahalanobis distance criterion (\citealt{Xiang2008}, see Appendix~\ref{app:overlap_robustness} and Table~\ref{tab:overlap_robustness}). Within the overlap-restricted sample, 97 systems are associated with cluster~A1, 72 with A2, 573 with B, and 101 with C.

The resulting distributions of formation-related parameters for the mapped synthetic systems are shown in Figure~\ref{fig12}. The three parameters considered, the gas availability parameter $\mathcal{G}_1$, the gas mass fraction $f_{\mathrm{gas}}$, and the ice--rock mass ratio, are defined in Section~\ref{method}. For clarity, the clusters are referred to using descriptive category labels motivated by their observational properties (Figures~\ref{fig4}–\ref{fig8}). Cluster~C corresponds to the very-massive gas giant (VMGG) group, cluster~A2 to a WJ--dominated (WJD) group, cluster~A1 to the hot-giant (HG) group, and cluster~B to the lower-mass giant (LMG) group. These labels are adopted for descriptive purposes only and are used below to summarise the trends in Figure~\ref{fig12}.

In terms of the gas availability parameter $\mathcal{G}_1$, the mapped synthetic populations associated with all clusters are concentrated at relatively high values, consistent with formation occurring while sufficient disk gas remained available for envelope accretion. The VMGG group occupies the highest $\mathcal{G}_1$ regime overall (median $\mathcal{G}_1 \approx 0.864$), indicating a stronger association with formation under particularly gas-rich disk conditions. The WJD group also exhibits systematically elevated $\mathcal{G}_1$ values relative to the HG group, with median values of $\approx 0.717$ and $\approx 0.667$, respectively. This suggests that WJD systems tend to form, on average, in slightly more gas-rich environments than HG systems. The LMG group spans a broader range of $\mathcal{G}_1$ values while maintaining a median ($\approx 0.708$) comparable to that of the WJD group, indicating that lower-mass giant planets can form across a wider range of gas-availability conditions.

The distributions of $f_{\mathrm{gas}}$ further differentiate the clusters. Systems associated with the HG, WJD, and VMGG groups are overwhelmingly gas-dominated, with $f_{\mathrm{gas}}$ exceeding 90\% in most cases. In contrast, the LMG group spans a broad range of gas mass fractions, extending from gas-poor systems to planets with substantial gaseous envelopes. A similar distinction is evident in the ice--rock mass ratio: the HG, WJD, and VMGG groups are strongly concentrated toward high values, whereas the LMG group exhibits a broader distribution extending toward lower ice--rock ratios. Although a subset of LMG systems reaches high ice--rock values, the overall distribution is shifted toward lower values compared to the other clusters, and spans a wider range, indicating greater diversity in solid composition.

To place these distributions in an orbital context, the mapped synthetic systems are next projected into the logarithmic $a$--$q_{\mathrm{p/s}}$ plane, as shown in Figure~\ref{fig13}. In this representation, the formation-related parameters do not vary in a simple or monotonic manner with either orbital distance or planet--star mass ratio across all clusters. For $\mathcal{G}_1$ (Figure~\ref{fig13}, top panel), the LMG group shows a trend in which systems at smaller final semi-major axes tend to be associated with higher $\mathcal{G}_1$ values. Since $\mathcal{G}_1$ traces the amount of disk gas available at the onset of formation, this suggests that these systems formed, on average, earlier in more gas-rich disks. Because the semi-major axes shown here are final post-migration values, this trend should not be interpreted as a direct formation-radius sequence. No comparable trend is observed for the VMGG group, which is instead concentrated at high $q_{\mathrm{p/s}}$ values and is characterised by systematically high $\mathcal{G}_1$ across its parameter space. The HG group occupies a relatively restricted region of the plane and does not exhibit a pronounced internal trend in $\mathcal{G}_1$. In particular, the HG systems show a mixed distribution, with no clear monotonic dependence on $q_{\mathrm{p/s}}$. In contrast, the WJD group shows a tendency for systems at larger semi-major axes and higher $q_{\mathrm{p/s}}$ values to be associated with higher $\mathcal{G}_1$, suggesting that the most massive WJD systems preferentially form under relatively gas-rich disk conditions.

The distributions of $f_{\mathrm{gas}}$ and the ice--rock mass ratio in the $a$--$q_{\mathrm{p/s}}$ plane (Figure~\ref{fig13}, middle and bottom panels) further highlight cluster-dependent behaviour. Within the LMG group, both quantities increase systematically with $q_{\mathrm{p/s}}$, such that more massive systems are associated with higher gas fractions and higher ice--rock ratios. In contrast, the HG, WJD, and VMGG groups remain concentrated at high values of both parameters over a relatively narrow range of $q_{\mathrm{p/s}}$, with no strong visual dependence on either $q_{\mathrm{p/s}}$ or $a$. These cluster-dependent trends are statistically supported by non-parametric tests applied to the mapped synthetic population (Appendix~\ref{stat_diag_3D_SP}).

\section{Discussion}\label{discuss}

\subsection{Interpretation of the two-stage GMM clustering}

The relative contribution of each input parameter to the final five-cluster solution is quantified using ANOVA F-values (Figure~\ref{fig3}c). The highest F-values are obtained for $R_\mathrm{Hill}$, $R_\mathrm{p}$, and $P_{\mathrm{orb}}$, indicating that these parameters play a dominant role in separating the clusters in the adopted feature space. In contrast, $N_\mathrm{p}$ and $\log g_\mathrm{p}$ yield the smallest F-values, suggesting that they contribute more weakly to the global discrimination between clusters.

Following the second-stage subdivision of cluster~A, the refined partition shows increased overlap between neighbouring components, reflecting the resolution of finer distinctions within a population that is already internally differentiated (see Section~\ref{GMM_two_stage_results}). Such behaviour is expected in a hierarchical clustering framework, where subdividing a dominant group naturally leads to less sharply separated sub-components. Rather than indicating a loss of clustering quality, this increased overlap signals that the second-stage GMM is capturing more subtle internal differentiation within an otherwise compact population.

The maximum posterior cluster probability for most planets exceeds 0.8, indicating that individual objects are generally well associated with a dominant mixture component (Figure~\ref{fig3}b). In contrast, cluster assignment stability, estimated via bootstrap-based stability values and typically distributed in the range 0.2--0.4, reflects the probabilistic nature of the GMM, particularly in the second-stage clustering where the boundary between sub-clusters A1 and A2 is intrinsically fuzzy. Objects with moderate stability values are therefore best understood as occupying transitional regions near cluster interfaces, rather than as exhibiting unstable or poorly constrained classifications. Such behaviour is expected when modelling continuous physical populations with gradual evolutionary differences.

Viewed in this light, the partial overlap apparent in the low-dimensional visualisation of the final clustering (Figure~\ref{fig3}d) should be interpreted as a faithful projection of the original parameter space, rather than as an artefact of the dimensionality-reduction technique.

\subsection{Physical validation of the A1--A2 subdivision}

The second-stage subdivision of cluster~A into A1 and A2 refines a previously dominant population and is therefore accompanied by increased overlap in global clustering diagnostics. It is therefore useful to examine whether this subdivision is consistent with a physically meaningful separation rather than being driven purely by statistical effects of the clustering procedure.

Although the normalised parameter profiles of A1 and A2 differ only modestly, the second-stage GMM produces a systematic redistribution of planet categories, with warm Jupiters and a subset of hot Jupiters preferentially assigned to A2. The subdivision is primarily controlled by derived dynamical quantities, notably $R_{\mathrm{Hill}}$, $v_{\mathrm{esc}}/v_{\mathrm{orb}}$, and $\Theta_{\mathrm{Saf}}$, which combine planetary bulk properties with orbital parameters. In contrast, direct size measures such as $R_\mathrm{p}$ play a minor role. This suggests that the second-stage GMM isolates differences in the dynamical configuration of the planet–star system rather than simply re-partitioning planets by size.

Importantly, the A1--A2 distinction also appears in parameter planes not directly used in the clustering analysis. In the $T^*_\mathrm{eff}$--$T_\mathrm{eq}$ plane (Figure~\ref{Append2}), planets assigned to A2 systematically occupy lower $T_\mathrm{eq}$ values than those in A1 at comparable stellar temperatures, particularly at lower $T^*_\mathrm{eff}$. This offset suggests weaker irradiation and larger orbital separations for A2 systems, consistent with their association with warm Jupiters. 

While $T_\mathrm{eq}$ and stellar properties are correlated with some of the dynamical parameters entering the clustering (e.g. $R_{\mathrm{Hill}}$ and $v_{\mathrm{esc}}/v_{\mathrm{orb}}$), the persistence of the A1--A2 separation in this physically interpretable projection supports the interpretation of the second-stage GMM subdivision as reflecting a meaningful differentiation within the gas-giant population.

\subsection{Transitional nature of hot Saturns and low-confidence cases}

The low-confidence cases identified in the cluster mapping analysis (Figure~\ref{fig7}) provide additional insight into the continuity of the underlying planet population. A small subset of planets exhibits low-confidence cluster assignments, characterised by comparable posterior probabilities for more than one cluster. These objects are predominantly associated with the HS population, which occupies an intermediate region of parameter space between clusters~A and~B.

This behaviour is consistent with recent studies indicating that HSs span a wide range of formation and evolutionary pathways~\citep{Xie2025}, bridging regimes associated with both lower-mass gas-rich planets and more massive gas giants. Their observed diversity in bulk density and radius inflation~\citep{Bouchy2024} naturally leads to overlap with multiple clusters in an unsupervised framework. The comparison between individual parameter profiles and the median profiles of both the primary and secondary clusters highlights the intrinsically ambiguous nature of these systems, reflecting genuine physical intermediacy rather than shortcomings of the clustering procedure.

\subsection{Linking observational clusters to formation pathways}

Mapping the observationally defined GMM clusters onto a set of synthetic populations derived from planet formation simulations enables a physically motivated interpretation of the clustering results. The synthetic population spans most of the parameter space occupied by clusters HG (A1), LMG (B), and VMGG (C), with the distribution in $R_\mathrm{p}$ reflecting the use of mass--radius relations to infer planetary radii for the synthetic populations.

The formation-related parameters provide complementary constraints on the physical origin of the clusters. The $\mathcal{G}_1$ traces formation timing relative to disk dispersal, the  $f_{\mathrm{gas}}$ reflects envelope accretion efficiency, and the ice--rock mass ratio encodes the composition of the solid component and hence the formation region. Differences in both the distributions and internal correlations of these parameters across clusters suggest that the GMM-defined groups correspond to systematic variations in formation epoch, gas accretion history, and solid growth pathways, rather than arising solely from statistical partitioning of the data by the clustering algorithm.

A particularly informative comparison is provided by the VMGG and HG clusters. Although both groups are mainly dominated by UHJs, FHJs, HJs, and occupy similar regions in orbital period and radius, they are cleanly separated at the first stage of the GMM classification. This separation occurs primarily along the planet--star mass ratio axis, with VMGG systems concentrated at higher masses, typically within $\log_{10}(q_{\mathrm{p/s}})\simeq-2.5$ to $-2.0$. Independent statistical diagnostics applied to the mapped synthetic population (Appendix~\ref{stat_diag_3D_SP}) further show that VMGG planets are characterised by systematically higher $\mathcal{G}_1$ values and a strong positive correlation between $\mathcal{G}_1$ and $f_{\mathrm{gas}}$, consistent with formation in gas-rich disks at relatively early epochs, where envelope growth is closely coupled to formation timing.

In contrast, the HG cluster exhibits weaker internal correlations involving $f_{\mathrm{gas}}$, despite sharing broadly similar orbital properties. This difference implies that, while VMGG and HG planets may converge to comparable post-migration configurations, their growth histories are not identical. In particular, the VMGG population appears to reflect a regime in which relatively early formation and efficient gas accretion jointly favour the emergence of very massive gas giants, whereas the HG population is characterised by a broader range of accretion efficiencies and, on average, later formation timescales.

The LMG (B) cluster exhibits a distinctive combination of broad parameter distributions and moderate internal correlations among the formation-related quantities. Compared to the other clusters, the $\mathcal{G}_1$ distribution spans a wider range and extends to substantially lower values ($\mathcal{G}_1 \sim 0.4$), although its median value remains comparable to that of the WJD (A2) group. This suggests that many LMG systems formed relatively later, under more diverse disk conditions. The $f_{\mathrm{gas}}$ remains systematically low across the cluster, while the Ice--Rock ratio exhibits a broad distribution ranging from highly ice-rich systems to substantially lower values. In addition, systems at smaller final semi-major axes tend to be associated with higher $\mathcal{G}_1$ values (Fig.~\ref{fig13}), qualitatively resembling radial trends expected in some inside-out formation scenarios~\citep{Chatterjee2014}, although such behaviour is not observed in the other clusters. In Table~\ref{tab:kendall_tau_clusters}, the moderate negative correlations of $\mathcal{G}_1$ with both the Ice--Rock ratio and $f_{\mathrm{gas}}$ indicate that, within the LMG population, systems associated with earlier formation epochs tend to exhibit both lower envelope fractions and lower ice enrichment. Together, these trends suggest that the LMG population does not correspond to a single dominant formation pathway, but instead reflects a broader diversity of formation environments and growth histories compared to the other clusters.

The WJD (A2) cluster occupies an intermediate regime between these pathways. Its strong correlation between $\mathcal{G}_1$ and the ice--rock mass ratio (0.86), together with a moderate correlation with $f_{\mathrm{gas}}$ (0.57), is consistent with coherent early growth in the outer disk, where both the solid core and gaseous envelope are built up efficiently prior to migration (see Table~\ref{tab:kendall_tau_clusters}). Within the adopted migration framework (Appendix~\ref{close_mig}), the WJD population is also characterised by systematically higher $\mathcal{G}_1$ values than the HG group, suggesting that these systems tend to form earlier, under more gas-rich disk conditions, while their subsequent orbital evolution does not bring them into the closest-in orbital regime. The dominance of WJs in this cluster therefore suggests that, although the initial growth conditions resemble those of massive gas giants, inward migration or orbital evolution is halted before the planets reach ultra-short-period orbits.

At the same time, the choice of the 3D mapping is not arbitrary. The adopted axes correspond to the dominant contributors to the first three principal components (PC1--PC3; Table~\ref{tab:pca_loadings}), and two of them rank among the highest ANOVA F-values in the first-stage GMM classification. The choice of this parameter space therefore has quantitative statistical support from both dimensionality-reduction and clustering diagnostics.

Although these trends are statistically supported, their physical interpretation remains subject to several modelling assumptions. In particular, the adopted late-stage orbital evolution assumes inward migration within $\sim$0.1~AU after formation (Appendix~\ref{close_mig}), which may affect the detailed orbital distribution near cluster boundaries. Planetary radii are subsequently inferred from mass--radius relations, and transit probabilities are applied to construct the observable synthetic population (Appendix~\ref{app:transit_probability}). More generally, this migration treatment should be regarded as a simplified mapping that relocates synthetic planets into the observational domain rather than a fully self-consistent model of migration physics. The present results therefore reflect the formation--migration connection under this prescription, while the same framework can be applied to synthetic populations generated with alternative migration scenarios to assess how different migration mechanisms influence the inferred links between observed clusters and planet formation pathways.

Crucially, the dominant cluster-level trends remain robust to the above uncertainties. In particular, the synthetic population consistently indicates that the VMGG cluster formed at systematically earlier epochs than both the HG and WJD clusters, as supported by its higher $\mathcal{G}_1$ values, strong $\mathcal{G}_1$--$f_{\mathrm{gas}}$ correlation, and large $q_\mathrm{p/s}$. While the relative ordering between HG and WJD in formation time may depend on details of radius evolution and late migration, the early formation of the VMGG population emerges as a stable outcome of the combined clustering and mapping analysis.

\section{Conclusions}\label{conclusion}

In this work, a subset of the observed exoplanet population is classified using a probabilistic, two-stage Gaussian mixture model (GMM). The clustering is performed in a feature space that includes several dynamical parameters, and the resulting groups are primarily separated by derived dynamical quantities, as indicated by their high ANOVA F-values. These quantities encode dynamical scale and interaction strength, namely the orbital period ($P_\mathrm{orb}$), planetary radius ($R_\mathrm{p}$), planetary surface gravity ($g_{\rm p}$), Hill radius ($R_\mathrm{Hill}$), Safronov number ($\Theta_\mathrm{Saf}$), planet–star mass ratio ($q_\mathrm{p/s}$), escape-to-orbital velocity ratio ($v_\mathrm{esc}/v_\mathrm{orb}$), semi-major axis in units of stellar radius ($a/R_\mathrm{s}$), and the number of confirmed planets in the system ($N_\mathrm{p}$), indicating that the classification is governed more by the system’s dynamical architecture than by directly observed properties alone. The second-stage refinement resolves additional differentiation within the dominant gas-giant population, naturally yielding partially overlapping sub-groups that are consistent with a continuous underlying population.

A quantitative link between the machine-learning-defined clusters and pebble-accreted synthetic populations is established through a statistically motivated three-dimensional mapping anchored by the dominant contributors to principal components PC1--PC3 (Table~\ref{tab:pca_loadings}). Despite modelling uncertainties, including the adopted late-migration prescription within $\sim$0.1~AU and mass--radius-based synthetic radii, the cluster-level trends remain stable and reveal distinct formation pathways. In particular, the very-massive gas-giant (VMGG) cluster is consistent with early formation in gas-rich disks, while the hot-giant (HG) cluster exhibits more heterogeneous accretion histories. Within the adopted migration framework, the WJ-dominated (WJD) cluster shows coherent links between formation timing, gas accretion, and ice enrichment, and is associated with systematically earlier formation than the HG population despite not migrating into the closest-in orbital regime. In contrast, the low-mass giant (LMG) cluster displays broader and partially anti-correlated trends, indicating more diverse growth histories. These results demonstrate that dynamically motivated machine-learning classifications can be connected to physically motivated synthetic populations in a statistically controlled manner, providing a data-driven framework for linking observed exoplanet demographics to planet formation pathways.

\section*{Data availability}

The data underlying this work are available from the NASA Exoplanet Archive at \url{https://exoplanetarchive.ipac.caltech.edu/cgi-bin/TblView/nph-tblView?app=ExoTbls&config=PS}. The data products associated with this work are available at \url{https://github.com/yiduann/gmm-exoplanet-clusters}.

\begin{acknowledgements}

This work was supported by the National Science and Technology Council of Taiwan (NSTC 114-2917-I-564-044, 114-2124-M-008-003, 114-2112-M-008-016), Taiwan Space Union (TSU), and by the Center for Star and Planet Formation at the Globe Institute, University of Copenhagen, Denmark. A.J. and H.S.W. acknowledge support from the Carlsberg Foundation through the FIRSTATMO project. H.J.H. acknowledges support from eSSENCE (eSSENCE@LU 9:3), the Swedish National Research Council (grant 2023-05307), the Crafoord Foundation, and the Royal Physiographic Society of Lund. We also sincerely thank the referee for the careful reading of the manuscript and the constructive comments and suggestions, which greatly improved the quality and clarity of this work.

\end{acknowledgements}

\bibliographystyle{aa} 
\bibliography{sn-bibliography}

\begin{appendix}

\section{Definitions of derived dynamical parameters}\label{app:dynamical_quantities}

Several dynamical quantities adopted as input features in the clustering analysis are not directly provided by the NASA Exoplanet Archive, but are instead derived from reported planetary and stellar properties. These parameters encode characteristic dynamical scales, interaction strengths, and scattering efficiencies of the planet--star system, and are therefore useful for distinguishing different dynamical regimes within the observed population.

The planet--star mass ratio is defined as
\begin{equation}
q_{\mathrm{p/s}} = \frac{M_{\mathrm{p}}}{M_{\mathrm{s}}},
\label{eq:mass_ratio}
\end{equation}
where $M_{\mathrm{p}}$ and $M_{\mathrm{s}}$ denote the planetary and stellar masses, respectively. This dimensionless quantity provides a natural measure of the dynamical importance of the planet relative to its host star and enters explicitly into several other derived parameters.

The orbital scale of each system is characterised using the scaled semi-major axis, defined as
\begin{equation}
\frac{a}{R_{\mathrm{s}}},
\label{eq:a_Rs}
\end{equation}
where $a$ is the orbital semi-major axis and $R_{\mathrm{s}}$ is the stellar radius. This parameter provides a dimensionless measure of orbital separation that is directly related to irradiation level and tidal interaction strength.

The Hill radius quantifies the gravitational sphere of influence of the planet and is given by
\begin{equation}
R_{\mathrm{Hill}} = a \left( \frac{M_{\mathrm{p}}}{3\,M_{\mathrm{s}}} \right)^{1/3},
\label{eq:R_Hill}
\end{equation}
where $a$ is the semi-major axis. In this work, $R_{\mathrm{Hill}}$ is expressed in astronomical units (AU) and serves as a characteristic dynamical length scale of the planet--star system.

The Safronov number provides a compact measure of the efficiency of gravitational scattering relative to orbital motion and is defined as
\begin{equation}
\Theta_{\mathrm{Saf}} 
= \frac{a}{R_{\mathrm{p}}}\frac{M_{\mathrm{p}}}{M_{\mathrm{s}}}.
\end{equation}
Larger values of $\Theta_{\mathrm{Saf}}$ correspond to systems in which gravitational encounters are more effective compared to the characteristic orbital velocity.

For convenience, we also consider the escape-to-orbital velocity ratio,
\begin{equation}
\frac{v_{\mathrm{esc}}}{v_{\mathrm{orb}}} = \sqrt{2\,\Theta_{\mathrm{Saf}}},
\end{equation}
which provides an equivalent measure of the relative strength of planetary gravity and orbital motion.

Finally, the planetary surface gravity is computed as
\begin{equation}
g_{\mathrm{p}} = \frac{G M_{\mathrm{p}}}{R_{\mathrm{p}}^2},
\end{equation}
and is reported in logarithmic form as $\log g_{\mathrm{p}}$, expressed in cgs units (cm\,s$^{-2}$). This quantity provides an additional measure of planetary compactness and gravitational binding strength.

Together, these derived parameters provide a physically motivated description of the dynamical environment of each planet and form the basis of the feature space used in the unsupervised clustering analysis.

\section{t-SNE visualisation of the parameter space}\label{tSNE_append}

As an exploratory step, the reliability of the t-SNE (t-distributed Stochastic Neighbor Embedding) embedding was assessed using a set of complementary diagnostics. The trustworthiness of the embedding is consistently high, ranging from 0.983 to 0.995 for neighbourhood sizes $k=5$--30, indicating good preservation of local neighbourhood relationships. The neighbourhood overlap between the original space and the embedding lies between 0.64--0.78 over the same range of $k$, indicating that the majority of original nearest neighbours are retained and that the local geometry is largely preserved.

Clustering quality indices computed in the original feature space yield a SIL score of 0.316, a DBI of 1.316, and a CHI of 191.3, indicative of moderate cluster separation and compactness. When evaluated in the t-SNE space, the corresponding values (SIL=0.305, DBI=1.406, CHI=325.1) remain comparable, with the increase in the CHI reflecting improved visual separation relative to cluster compactness. A permutation test on the SIL score further yields an observed value of 0.305 compared to a null expectation of $-0.085 \pm 0.032$, corresponding to a $p$-value of 0.005, indicating that the apparent separation is statistically significant and unlikely to arise by chance.

The robustness of the embedding across different perplexity values is supported by Procrustes disparity values in the range 0.136--0.242, demonstrating stable global relationships under changes in the embedding parameters. In addition, a linear classifier trained on the 2D t-SNE coordinates achieves a cross-validated accuracy of $0.896 \pm 0.018$ and an F1 score of $0.714 \pm 0.053$, indicating that the main classes remain reasonably well separated in the low-dimensional projection. These performance values are not used as a primary classification benchmark, but rather as a sanity check that the t-SNE embedding preserves sufficient clustering information to inform the subsequent GMM-based clustering analysis.

\section{Diagnostic analysis of input features}\label{input_append}

A series of complementary diagnostic analyses was performed to characterise the distributions, redundancy, and stability of the adopted orbital and physical parameters, and to assess their suitability as inputs for the GMM clustering.

\begin{figure}
    \centering
    \includegraphics[width=0.49\textwidth]{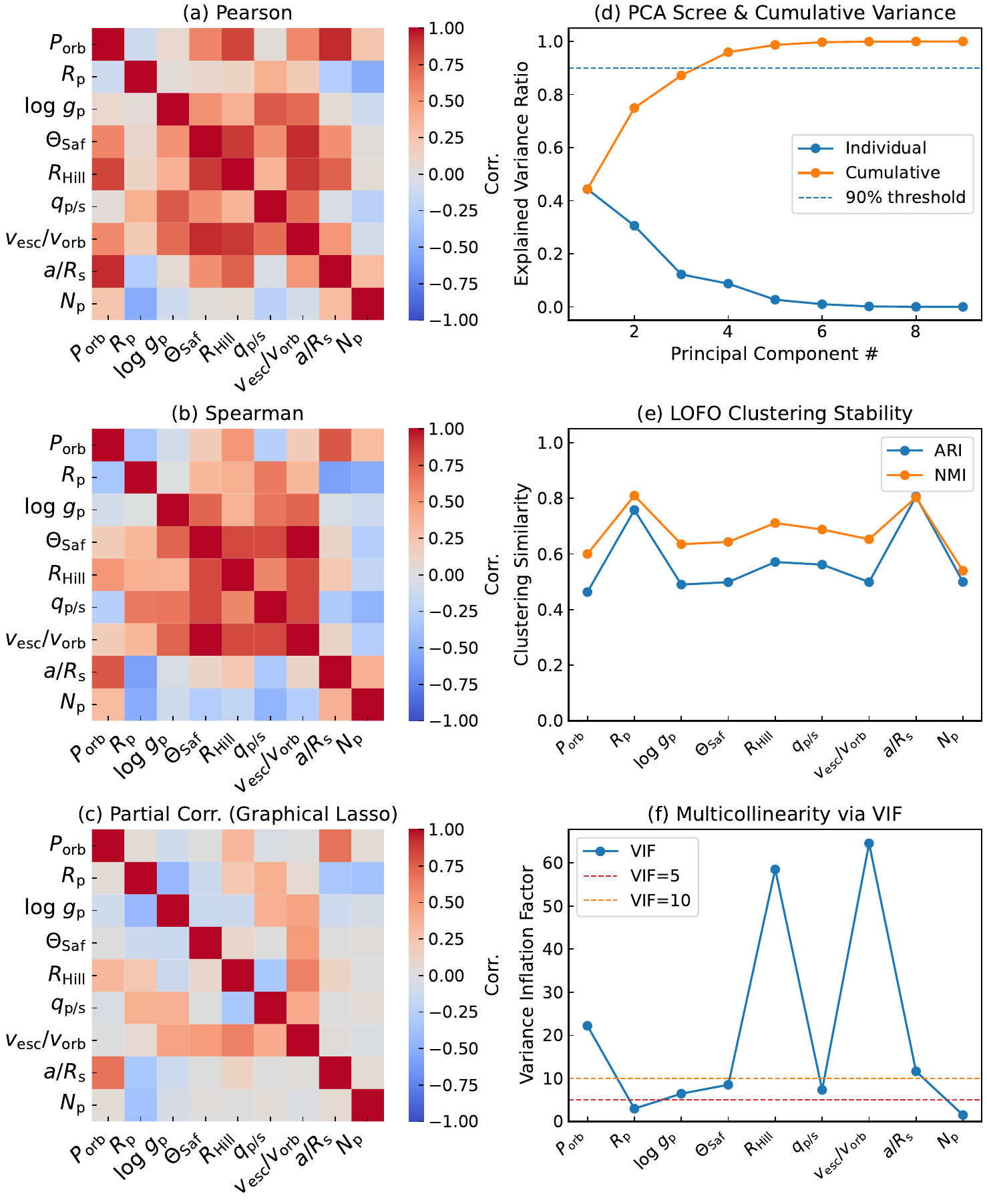}
    \caption{Diagnostic analysis of parameter dependencies, dimensionality, and clustering robustness for the standardised input feature set. (a) Pearson correlation matrix of the adopted orbital and physical parameters, quantifying linear pairwise correlations. (b) Spearman rank correlation matrix, illustrating monotonic relationships that may not be strictly linear. (c) Partial correlation matrix estimated via Graphical Lasso, highlighting direct conditional dependencies between parameters after accounting for the influence of all others. (d) Principal Component Analysis (PCA) scree plot showing the explained variance ratio of individual components (blue line) and the cumulative explained variance (orange line); the dashed horizontal line marks the 90\% cumulative variance threshold. (e) Leave-one-feature-out (LOFO) clustering stability test, in which each parameter is removed in turn and the resulting GMM clustering is compared to the baseline solution using the adjusted Rand index (ARI) and the normalised mutual information (NMI), providing a measure of the sensitivity of the clustering to individual features. (f) Variance inflation factor (VIF) for each parameter, used to assess multicollinearity within the feature set; dashed horizontal lines indicate commonly adopted thresholds at VIF = 5 and 10.}\label{Append1}
\end{figure}

The pairwise correlation analysis reveals a structured pattern of dependencies among the input parameters (Figure~\ref{Append1}). In both the Pearson and Spearman matrices (Figure~\ref{Append1}a,b), strong correlations are observed between quantities that are either physically related by construction or encode similar dynamical information. In particular, $\Theta_\mathrm{Saf}$ exhibits an almost one-to-one correspondence with $v_\mathrm{esc}/v_\mathrm{orb}$, and strong correlations with both $R_\mathrm{Hill}$ and $q_\mathrm{p/s}$, reflecting their shared dependence on planetary mass, radius, and orbital distance proxies encoded by $P_{\mathrm{orb}}$ and $a/R_\mathrm{s}$. Likewise, the pronounced correlation between $P_{\mathrm{orb}}$ and $a/R_\mathrm{s}$ arises naturally from their common Keplerian scaling. Rather than motivating feature removal, these correlations indicate that multiple parameters provide overlapping but physically meaningful descriptions of a small number of fundamental dynamical scales.

To distinguish direct dependencies from correlations induced by shared underlying variables, we compute the partial correlation matrix using Graphical Lasso (Figure~\ref{Append1}c). This analysis isolates a more limited set of conditional relationships after controlling for all other parameters, yielding a sparse representation of the dominant conditional dependencies. In this framework, $P_{\mathrm{orb}}$ retains a strong conditional link with both $a/R_\mathrm{s}$ and $R_\mathrm{Hill}$, while $R_\mathrm{Hill}$ remains directly connected to $v_\mathrm{esc}/v_\mathrm{orb}$ and $q_\mathrm{p/s}$, with opposite correlation signs. In contrast, many correlations that appear strong in the pairwise analysis are substantially weakened, indicating that they are largely mediated by a small number of underlying physical parameters rather than representing independent relationships.

The global variance structure of the feature space is further illustrated by the principal component analysis (PCA; Figure~\ref{Append1}d). The PCA defines a set of orthogonal axes (PC1--PC4) that successively capture the dominant variance while reducing redundancy among correlated variables. Although the first few components already account for more than 90\% of the total variance, the variance is distributed across multiple components rather than collapsing onto a single dominant mode, indicating that the parameter space retains several complementary dimensions. As summarised in Table~\ref{tab:pca_loadings}, PC1 is primarily driven by planetary scale-related parameters such as $R_\mathrm{p}$ and $q_\mathrm{p/s}$, while $R_\mathrm{Hill}$ emerges most prominently in PC3, reflecting variance associated with combined mass--distance effects after the dominant planetary-scale and orbital-distance contributions have been removed by the preceding components. The appearance of $a/R_\mathrm{s}$ in both PC2 and PC3 further highlights the multi-faceted role of orbital distance, whereas the presence of $R_\mathrm{p}$ across PC1--PC3 underscores its central role in linking planetary physical properties, orbital properties, and dynamical scales.

\begin{table}
\caption{Principal component (PC) loadings and physical interpretation for the PCA applied to the GMM input feature space.}
\label{tab:pca_loadings}
\renewcommand{\arraystretch}{1.2}
$$
\begin{array}{lc p{0.45\columnwidth}}
\hline
\noalign{\smallskip}
\text{PC} & \text{Explained variance} & \text{Main loadings} \\
\hline
PC1 & 0.444\ (44\%) & $R_\mathrm{p},\ q_\mathrm{p/s}, g_\mathrm{p}$ \\
PC2 & 0.306\ (31\%) & $R_\mathrm{p},\ g_\mathrm{p},\ N_\mathrm{p},\ a/R_\mathrm{s}$ \\
PC3 & 0.123\ (12\%) & $a/R_\mathrm{s},\ N_\mathrm{p},\ R_\mathrm{p},\ R_\mathrm{Hill}$ \\
PC4 & 0.088\ (9\%) & $N_\mathrm{p},\ a/R_\mathrm{s}$ \\
\hline
\end{array}
$$
\end{table}

The robustness of the clustering to individual parameters is assessed through a leave-one-feature-out (LOFO) analysis (Figure~\ref{Append1}e), in which each reduced clustering is compared to the baseline using ARI and NMI. In most cases, NMI remains relatively high, indicating that the overall cluster assignments are largely preserved, while variations in ARI mainly reflect changes in object-level assignments near cluster boundaries. The removal of $R_\mathrm{p}$ or $a/R_\mathrm{s}$ yields the highest ARI and NMI values, demonstrating that their information content is strongly redundant and largely recoverable from other correlated features. Conversely, removing $N_\mathrm{p}$ leads to the lowest NMI and a marked reduction in ARI, highlighting its comparatively independent contribution to stabilising cluster membership.

Finally, the degree of multicollinearity among the input parameters is quantified using the variance inflation factor (VIF; Figure~\ref{Append1}f). Several parameters exhibit elevated VIF values, most notably $v_\mathrm{esc}/v_\mathrm{orb}$, $R_\mathrm{Hill}$, and $P_\mathrm{orb}$, reflecting their shared dependence on planetary mass and orbital distance. While such multicollinearity would be problematic in a regression context, it is not detrimental in an unsupervised clustering framework such as the GMM, where the objective is to model the joint distribution of the data rather than to estimate independent parameter effects. In this setting, correlated dynamical descriptors encode physically coupled scales and collectively define the covariance structure that is explicitly learned by the GMM. Consistent with this interpretation, the LOFO analysis shows that the clustering assignments are not driven by any single high-VIF parameter, indicating that the associated information is distributed across correlated features rather than concentrated in individual variables.

Overall, the diagnostic analyses consistently indicate that the adopted parameters are not redundant observables, but complementary projections of a small number of physically coupled dynamical scales. The correlation and partial correlation analyses show that many strong pairwise relationships arise from shared underlying dependencies rather than direct coupling, while the PCA demonstrates that correlated and high-VIF parameters contribute to distinct orthogonal components rather than collapsing into a single dominant mode. Together with the LOFO and VIF results, this coherence shows that retaining correlated parameters does not merely duplicate information, but enables the GMM to capture the full distribution of the planetary parameter space, justifying the use of the complete feature set without aggressive parameter removal.

\section{Independent physical validation of the A1–A2 subdivision}\label{val_gmm_2nd}

\begin{figure}
\centering \includegraphics[width=0.49\textwidth]{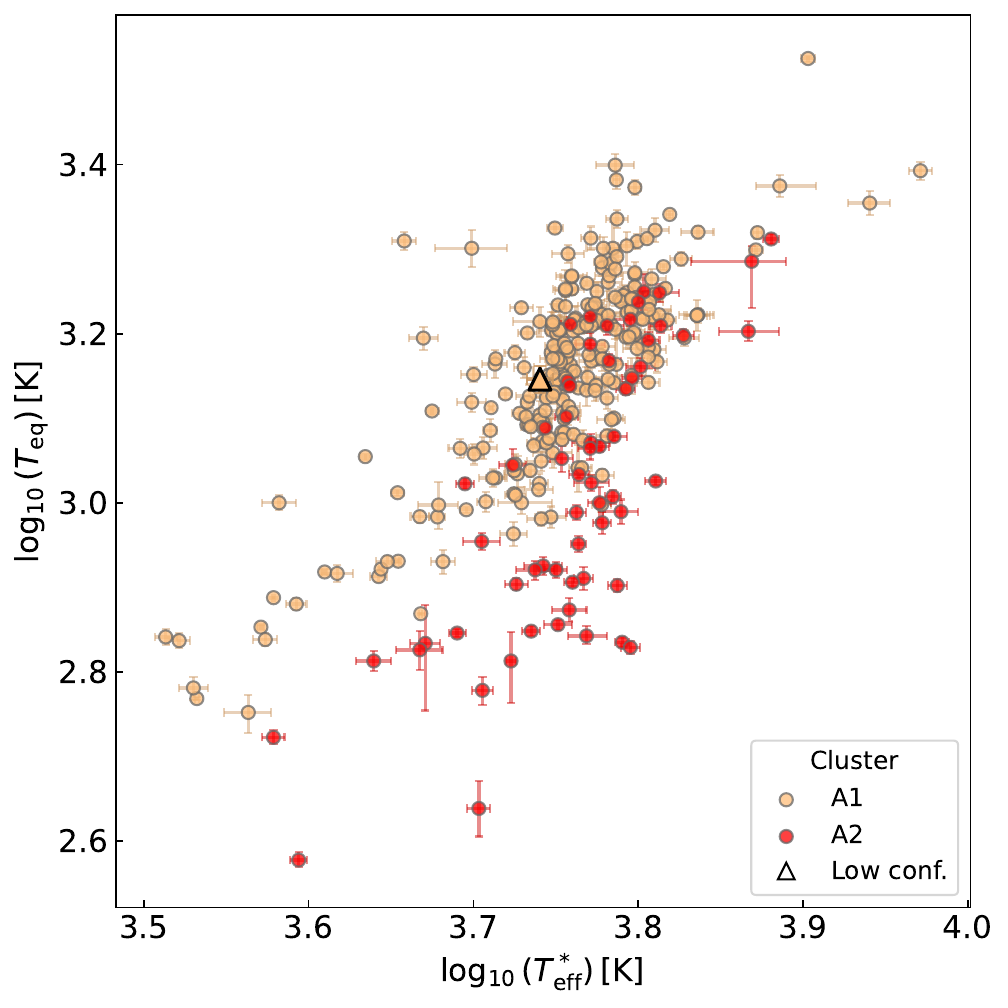} 
\caption{Distribution of A1 and A2 planets in the logarithmic plane of $T^*_\mathrm{eff}$--$T_\mathrm{eq}$. Triangles mark planets with low clustering confidence. }
\label{Append2}
\end{figure}

Figure~\ref{Append2} provides an independent physical validation of the second-stage GMM subdivision of cluster~A into A1 and A2. 
Because the separation between these two sub-clusters is relatively subtle in the GMM input space, and is accompanied by a decrease in global clustering metrics, it is important to verify whether the split corresponds to physically distinct populations rather than to a purely statistical artefact.

To this end, the distributions of A1 and A2 planets are examined in the plane of stellar effective temperature ($T^*_\mathrm{eff}$) and planetary equilibrium temperature ($T_\mathrm{eq}$), neither of which enters the GMM input set. 
This plane therefore provides an observationally independent projection of the clustered sample. 
Both sub-clusters follow the expected irradiation-driven trend of increasing $T_\mathrm{eq}$ with $T^*_\mathrm{eff}$. 
However, a systematic offset is observed at lower $T^*_\mathrm{eff}$, where planets assigned to A2 preferentially occupy lower equilibrium temperatures than those in A1 at comparable stellar temperatures.

This behaviour indicates that the A1–A2 subdivision is not merely a re-partitioning within the original feature space, but reflects a physically meaningful separation that persists in parameters external to the clustering. 
The result is particularly relevant given that cluster~A2 is dominated by warm Jupiters (Figure~\ref{fig6}), allowing differences in the stellar–planetary temperature relation to be assessed in an independent physical context.

\section{Synthetic population}\label{SP_append}

\subsection{Close-in migration}\label{close_mig}

In the synthetic population adopted here, planet formation and early disk-driven migration are terminated at the inner edge of the protoplanetary disk, such that planets remain at orbital distances of order $\sim$0.1~AU. This simplified setup ensures computational consistency and does not include additional post-formation migration channels, such as high-eccentricity migration driven by dynamical instabilities or secular interactions. As a consequence, the raw synthetic population does not fully populate the close-in orbital region sampled by the observed transiting giant planets. To enable a meaningful comparison with observations, we therefore apply an idealised post-formation inward migration step that relocates synthetic planets into the close-in regime in a controlled and time-limited manner. This procedure is intended as a mapping to observationally accessible orbits rather than as a detailed model of migration physics, while preserving the relative formation histories encoded in the synthetic population.

Post-formation orbital migration is simply simulated with the orbital evolution integrated via a fourth-order Runge--Kutta (RK4) scheme. Within this framework, the migration efficiency is assumed to increase with planet mass and to be regulated by disk depletion, without explicitly modelling gap opening or disk viscosity. All planets initially located within 1~AU are assumed to be present at the onset of the migration phase. Their subsequent orbital evolution is governed by a time-dependent inward migration law applied over a finite migration window. The evolution of the semi-major axis is governed by
\begin{equation}
\frac{da}{dt} = -\frac{a}{\tau(a,q)} \exp\!\left(-\frac{t}{\tau_{\rm dep}}\right),
\end{equation}
where $\tau(a,q)$ is an effective migration timescale that depends on the instantaneous semi-major axis $a$ and the planet--to--star mass ratio $q$, and $\tau_{\rm dep}$ denotes the disk depletion timescale. A value of $\tau_{\rm dep}=0.6\,\mathrm{Myr}$ was adopted to introduce a gradual temporal decay in migration efficiency. The migration equation was integrated over a finite interval $t \leq t_{\rm disk}$, with $t_{\rm disk}=0.15\,\mathrm{Myr}$ chosen to represent an early phase during which disk-driven migration is expected to operate efficiently. The migration timescale is parametrised as
\begin{equation}
\tau(a,q) = \tau_0
\left(\frac{q_{\rm ref}}{q_{\rm eff}}\right)^{\beta_q}
\left[ 0.05 + f(a)^{\eta_r} \right],
\end{equation}
with
\begin{equation}
f(a) = \frac{a - a_{\rm inner}}{a_{\rm out} - a_{\rm inner}},
\end{equation}
where $a_{\rm inner}$ denotes the inner disk truncation radius and $a_{\rm out}=1$~AU is the outer boundary of the modelled population. The baseline migration timescale was set to $\tau_0 = 0.5\,\mathrm{Myr}$, providing a moderate migration strength appropriate for giant planets during the adopted migration window. The exponent $\eta_r>1$ steepens the radial dependence of the migration timescale toward larger orbital radii, suppressing highly coherent inward migration of low-mass planets, while the exponent $\beta_q\simeq0.4$ parametrises the descriptive dependence of the migration timescale on the planet--to--star mass ratio. The additive constant 0.05 acts as a numerical regularisation near the inner boundary and does not affect the qualitative migration behaviour at larger orbital radii.

To restrict the prescription to the regime where it remains qualitatively applicable, the effective mass ratio was bounded within $3\times10^{-4} \le q_{\rm eff} \le 3\times10^{-3}$, corresponding approximately to Saturn- to a few-Jupiter-mass planets around solar-mass stars. $q_{\rm ref}\simeq10^{-3}$ anchors the migration efficiency to a Jupiter-mass planet. The orbital evolution was integrated from $t=0$ to $t=t_{\rm disk}$ using an RK4 scheme with 300 uniform time steps; a fixed step size was adopted for stable integration. Planets were prevented from migrating interior to the inner disk edge $a_{\rm inner}\simeq0.015$ AU.

\begin{figure}
    \centering
    \includegraphics[width=0.49\textwidth]{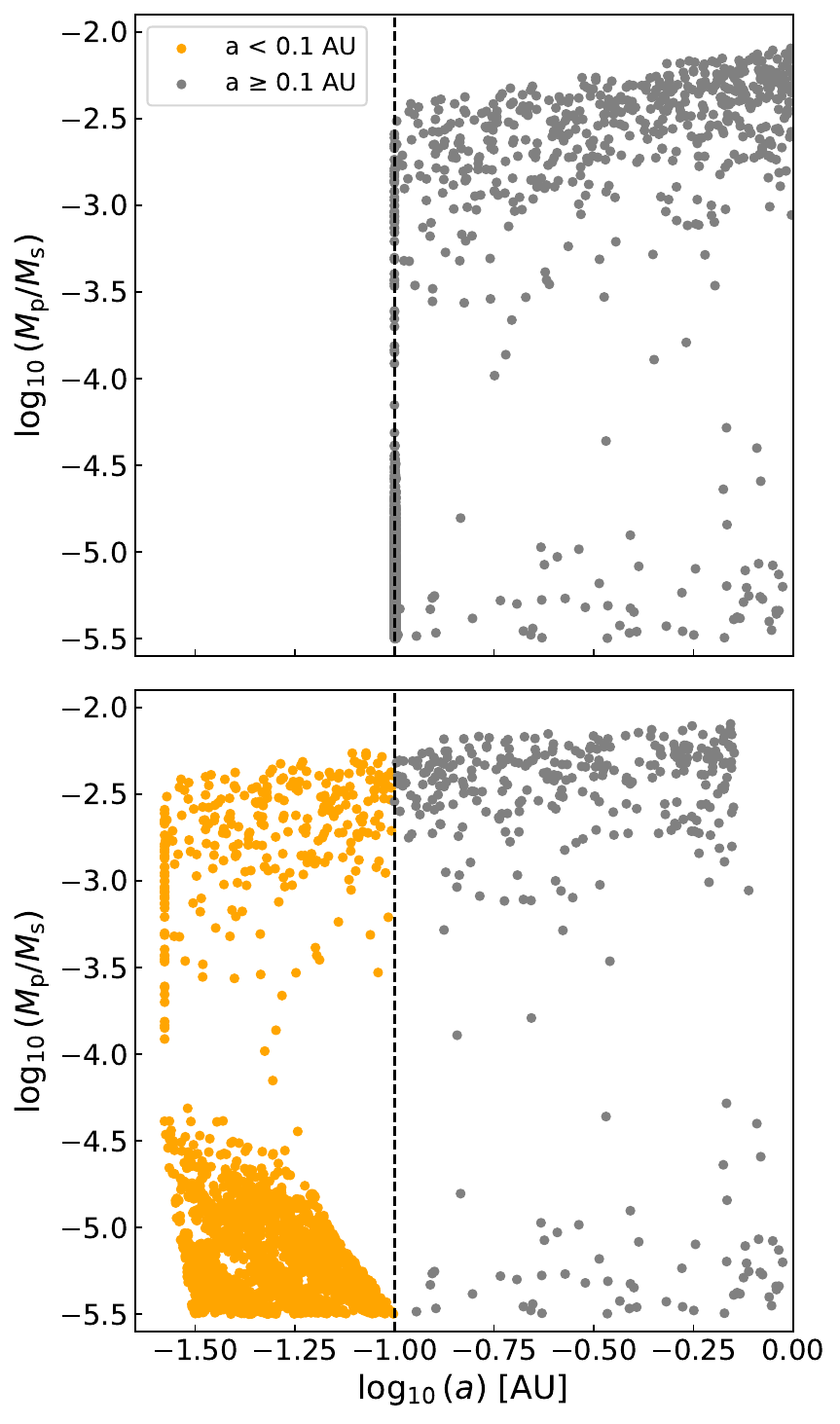}
    \caption{Distribution of planet--to--star mass ratio as a function of semi-major axis before (top) and after (bottom) applying the RK4 migration model. The dashed vertical line marks $a = 0.1$~AU, separating close-in planets (orange) from the outer population (gray).
    }\label{Append3}
\end{figure}

For very low-mass planets, defined here by $\log_{10} q<-4$, a piecewise treatment was adopted to avoid artificial pile-up at the inner disk edge. In this toy model, planets beyond $a \gtrsim 0.1~\mathrm{AU}$ were held fixed, whereas those already interior to this radius were allowed to migrate with rates further modulated by radial position and mass ratio. This treatment reduces excessive inward transport in the low-mass regime and yields a smoother final orbital distribution, as illustrated in Fig.~\ref{Append3}.

This migration prescription is not intended as a detailed physical model of disk--planet interactions. Instead, it serves as a controlled descriptive framework to mitigate known artefacts of the initial synthetic populations and to produce orbital architectures that are broadly consistent with population-level trends observed in real exoplanet systems. In particular, the prescription reduces the artificial pile-up of planets near 0.1 AU of the synthetic populations and allows massive planets to populate close-in orbits, consistent with the observed abundance of massive gas giants. At the same time, it avoids forcing low-mass planets to converge to a single characteristic radius. This treatment yields a smoother and more realistic distribution of semi-major axes, ensuring that derived quantities such as the Hill radius are evaluated in a physically meaningful orbital context when mapping synthetic populations onto observed systems.

\subsection{Mass-radius relation and transit probability}
\label{app:transit_probability}

To connect the migrated synthetic populations to potential observational
signatures, we first assign planetary radii to the synthetic populations and
then estimate their geometric transit probabilities. Planetary radii
are computed assuming solar-type host stars, using the empirical broken
power-law mass--radius relation inferred by \citet{Muller2024}. For each
planet, the radius is estimated via Monte Carlo sampling of the reported
uncertainties in the mass--radius relation. We adopt the median of the resulting radius distribution as the central
estimate, with the 16th and
84th percentiles used to characterise the uncertainty.

Given the assigned radii, the geometric transit probability for a
planet on a circular orbit is computed as
\begin{equation}
p_{\rm trans} = \frac{R_\mathrm{s} + R_{\rm p}}{a},
\end{equation}
where $a$ is the post-migration semi-major axis.

\begin{figure}
    \centering
    \includegraphics[width=0.49\textwidth]{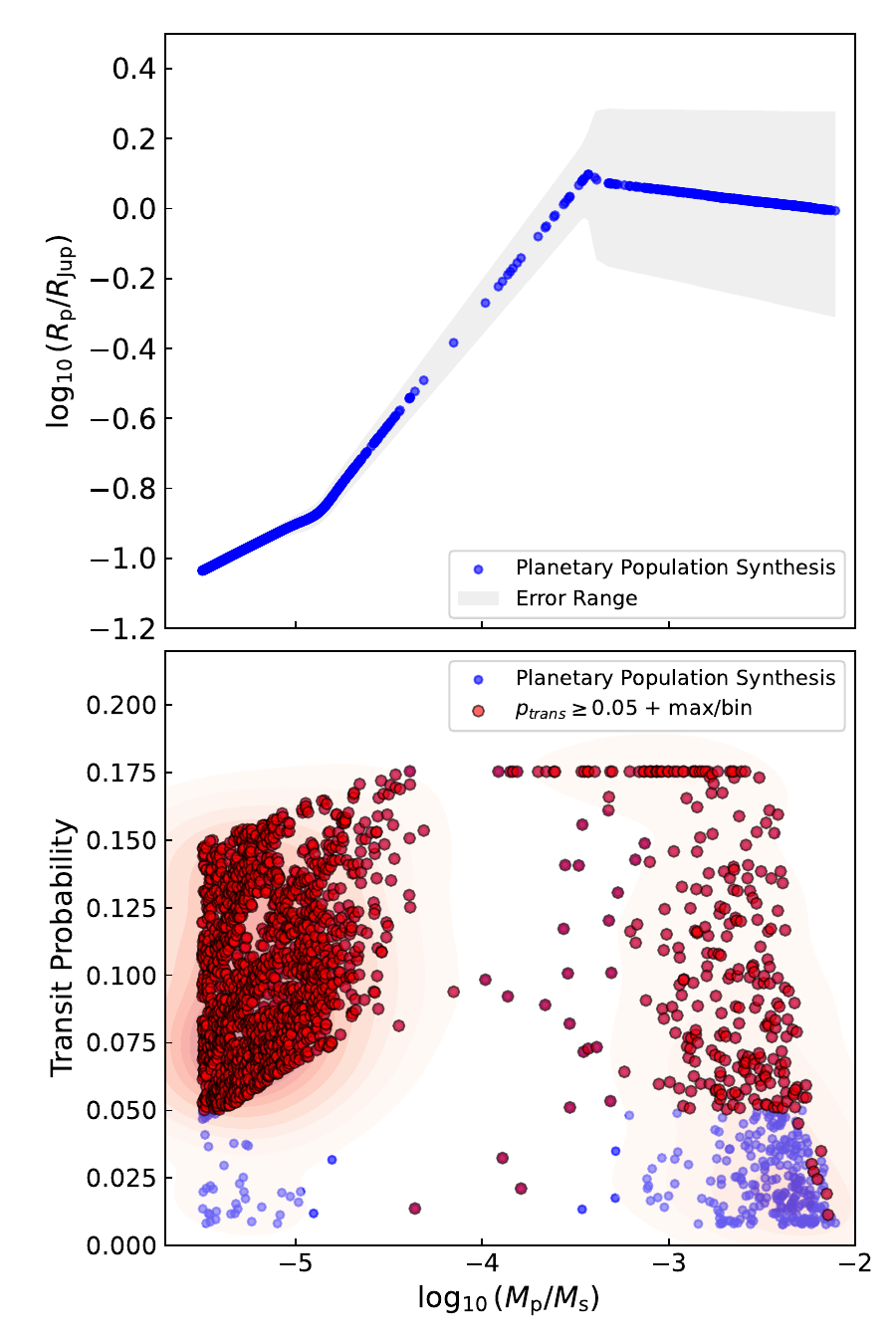}
    \caption{Distribution of planetary radii and transit probabilities derived from
the pebble accretion based synthetic populations. 
(Top) Predicted planet radius as a function of planet--to--star
mass ratio, based on a broken power-law mass--radius relation with
Monte Carlo sampling of the empirical uncertainties. Blue points denote
the median predicted radii, while the shaded region indicates the
$16$th--$84$th percentile range.
(Bottom) Transit probability as a function of
$\log_{10}(M_\mathrm{p}/M_\mathrm{s})$ for the full synthetic population
(blue points), with red contours showing the underlying density
distribution. Red points highlight a representative subsample selected
by retaining, in each mass bin, the planet with the highest transit
probability. This selection illustrates the envelope of detectability
expected from geometric transit bias.    }\label{Append4}
\end{figure}

Figure~\ref{Append4} illustrates the resulting planet radius and transit
probability distributions as a function of planet--to--star mass ratio for the synthetic populations. The top panel shows the predicted mass--radius relation with its associated uncertainties, while the bottom panel displays the corresponding transit probabilities. As
expected from geometric considerations, transit probability increases
toward smaller orbital separations and is therefore enhanced for close-in planets produced by inward migration.

To visualise the effective detectability envelope imposed by geometric transit bias, we further construct a representative subsample using a hybrid selection scheme. Within each mass-ratio bin, all planets with transit probabilities $p_{\rm trans} \geq 0.05$ are retained, while bins without such candidates retain only the planet with the highest transit probability. The 2190 selected synthetic populations highlighted in Fig.~\ref{Append4} therefore do not constitute a complete detectable sample, but instead trace the effective upper envelope associated with geometric transit selection. This procedure provides an intuitive illustration of how the underlying migrated population would be filtered by transit selection effects, while still preserving coverage across regions of parameter space where the transit probability remains intrinsically low.

\section{Robustness of the mapping to synthetic formation parameters}
\label{app:overlap_robustness}

To assess whether the synthetic planets used in the mapping are representative of the observed clusters, and to evaluate the impact of incomplete overlap between the observed and synthetic populations, we perform an additional robustness test in which only synthetic planets located within the multivariate domain of each observed cluster are retained. The overlap region is defined in the same three-dimensional feature space used for the KDE-based mapping ($\log R_{\mathrm{Hill}}$, $\log q_{\mathrm{p/s}}$, $\log R_{\mathrm{p}}$), using a Mahalanobis distance criterion derived from the observed cluster distributions~\citep{Xiang2008}. For a synthetic data point $\mathbf{x}$, the squared Mahalanobis distance is
\begin{equation}
d^2 = (\mathbf{x} - \boldsymbol{\mu})^\mathrm{T} \boldsymbol{\Sigma}^{-1} (\mathbf{x} - \boldsymbol{\mu}),
\end{equation}
where $\boldsymbol{\mu}$ and $\boldsymbol{\Sigma}$ are the location vector and covariance matrix of the observed cluster. The overlap region is defined by
\begin{equation}
d^2 \leq \chi^2_{k}(0.95),
\end{equation}
with $k=3$, corresponding to the 95\% confidence ellipsoid of the observed distribution.

The updated synthetic population substantially increases the number of mapped systems in clusters A1, A2, and C, enabling a more statistically robust assessment of the inferred formation properties. The formation-parameter distributions obtained from the overlap-restricted sample therefore provide a direct test of whether the inferred properties are sensitive to synthetic systems lying outside the observed parameter domain (Table~\ref{tab:overlap_robustness}).

For clusters A1, A2, and C, the overlap-restricted samples retain the large majority of the mapped synthetic planets ($N_o/N_f \gtrsim 95\%$). The median values remain nearly unchanged after applying the restriction, with median shifts ($\Delta$Med) that are negligible and well within the bootstrap confidence intervals. In addition, a leave-one-out (LOO) analysis is performed by recomputing the median after removing each synthetic planet in turn, yielding only minimal variation. Together, these results indicate that the inferred formation properties for these clusters are robust against incomplete overlap between the observed and synthetic populations and are not driven by individual data points.

For cluster B, the overlap-restricted analysis reveals substantial shifts in the median values, particularly in $f_{\mathrm{gas}}$ and the ice--rock mass ratio. This indicates that the synthetic population associated with cluster B extends significantly beyond the parameter region densely occupied by the observed systems. The LOO analysis nevertheless confirms that the overlap-restricted median values remain stable under the removal of individual planets, demonstrating that the inferred trends are not driven by a small number of outliers. We therefore adopt the overlap-restricted sample for the interpretation of cluster B in order to focus on synthetic systems that are most consistent with the observed parameter domain.

\begin{table}
\centering
\small
\caption{Robustness of formation-parameter inferences under an overlap-restricted synthetic sample.}
\label{tab:overlap_robustness}
\renewcommand{\arraystretch}{1.2}
\setlength{\tabcolsep}{3pt}
\begin{tabular}{@{}l l c c c c c@{}}
\hline
Cl. & Param & $N_o/N_f$ & Med$_f$/Med$_o$ & $\Delta$Med & LOO$_{\max}$ & 95\% CI \\
\hline
A1 & $\mathcal{G}_1$ & 97/101 & 0.675/0.667 & -0.008 & 0.671 & 0.65--0.68 \\
   & $f_{\rm gas}$   &        & 0.988/0.988 & 0.000 & 0.988 & 0.99--0.99 \\
   & Ice--Rock       &        & 1.000/1.000 & 0.000 & 1.000 & 1.00--1.00 \\
\hline
A2 & $\mathcal{G}_1$ & 72/74  & 0.716/0.717 & 0.001 & 0.718 & 0.70--0.73 \\
   & $f_{\rm gas}$   &        & 0.993/0.993 & 0.000 & 0.993 & 0.99--0.99 \\
   & Ice--Rock       &        & 1.000/1.000 & 0.000 & 1.000 & 1.00--1.00 \\
\hline
B  & $\mathcal{G}_1$ & 573/1908 & 0.684/0.708 & 0.024 & 0.708 & 0.69--0.72 \\
   & $f_{\rm gas}$   &          & 0.013/0.142 & 0.128 & 0.142 & 0.13--0.16 \\
   & Ice--Rock       &          & 0.449/0.971 & 0.522 & 0.972 & 0.96--0.98 \\
\hline
C  & $\mathcal{G}_1$ & 101/103 & 0.866/0.864 & -0.002 & 0.865 & 0.83--0.88 \\
   & $f_{\rm gas}$   &         & 0.995/0.994 & 0.000 & 0.995 & 0.99--0.99 \\
   & Ice--Rock       &         & 1.000/1.000 & 0.000 & 1.000 & 1.00--1.00 \\
\hline
\end{tabular}
\begin{tablenotes}
\footnotesize
\item Notes: Medians and bootstrap 95\% confidence intervals (CI) are reported for the overlap-restricted sample. Columns list the cluster label (Cl.), ratio of retained to total synthetic planets ($N_o/N_f$), median values for the full and overlap-restricted samples (Med$_f$/Med$_o$), median shift ($\Delta$Med), and the maximum median obtained from leave-one-out analysis (LOO${\max}$).
\end{tablenotes}
\end{table}

\section{Statistical diagnostics of the 3D-mapped synthetic populations}\label{stat_diag_3D_SP}

To assess whether the GMM-defined clusters identified in the 3D mapping correspond to statistically distinct populations, a set of non-parametric statistical diagnostics is applied to the synthetic populations (Table~\ref{tab:kruskal_tests}). In addition to formal significance testing, effect sizes are reported to quantify the magnitude of inter-cluster differences. For the Kruskal--Wallis tests, the effect size $\eta_\mathrm{H}^2$ is computed as $\eta_\mathrm{H}^2 = (H - k + 1)/(n - k)$, where $k=4$ is the number of clusters and $n=843$ is the total number of synthetic systems in the overlap-restricted sample, providing a normalised measure of separation between cluster distributions.

For all formation-related parameters considered, the Kruskal--Wallis statistics are large ($H \simeq 125$--$569$), with both asymptotic and permutation-based $p$-values $\lesssim 10^{-3}$, indicating highly significant differences across clusters. The largest effect sizes are obtained for $f_{\mathrm{gas}}$ and $q_{\mathrm{p/s}}$ ($\eta_\mathrm{H}^2 \simeq 0.67$), while $\mathcal{G}_1$ yields the weakest separation among the tested parameters, although its effect size ($\eta_\mathrm{H}^2 \simeq 0.15$) still falls within the conventionally defined large-effect regime~\citep{Tomczak2014,Borenson2023}.

\begin{table}
\caption{Kruskal--Wallis tests for synthetic population formation parameters across GMM clusters.}
\label{tab:kruskal_tests}
\centering
\renewcommand{\arraystretch}{1.2}
\begin{tabular}{lccc}
\hline
Parameter & $H$ statistic & $\eta_\mathrm{H}^2$ & $p_{\mathrm{asymp}}$ \\
\hline
$\mathcal{G}_1$     & 125.60 & 0.15 & $4.80\times10^{-27}$  \\
$f_{\mathrm{gas}}$  & 568.59 & 0.67 & $6.49\times10^{-123}$ \\
Ice--Rock           & 226.37 & 0.27 & $8.41\times10^{-49}$ \\
$q_{\mathrm{p/s}}$  & 569.08 & 0.67 & $5.09\times10^{-123}$ \\
$a$                 & 207.68 & 0.24 & $9.24\times10^{-45}$  \\
\hline
\end{tabular}
\begin{tablenotes}
\footnotesize
\item Notes: The $H$ statistic quantifies the overall separation between cluster distributions, while $p_{\mathrm{asymp}}$ denotes the corresponding asymptotic significance under the null hypothesis of identical distributions. The effect size $\eta_\mathrm{H}^2$ provides a normalised measure of the magnitude of inter-cluster differences. For all listed parameters, the permutation-based significance is highly robust, with $p_{\mathrm{perm}} = 3.3\times10^{-4}$.
\end{tablenotes}
\end{table}

To further quantify the contribution of individual formation parameters to cluster assignment, a multinomial logistic regression is applied to the 3D-mapped synthetic populations, and the resulting odds ratios are summarised in Table~\ref{tab:logit_odds}. Odds ratios greater than unity indicate an increased likelihood of assignment to a given cluster per $+1\sigma$ change in the corresponding standardised parameter, while values below unity indicate a decreased likelihood. The largest odds ratio is obtained for $q_{\mathrm{p/s}}$ in the VMGG cluster, consistent with the association of this group with high planet--star mass ratios. For the same cluster, $\mathcal{G}_1$ also exhibits an elevated odds ratio, exceeding that of the WJD and HG cluster.

In the HG cluster, the strongest positive association is found for $f_{\mathrm{gas}}$, whereas the WJD cluster shows its largest odds ratios for $a$ and $q_{\mathrm{p/s}}$. In contrast, the LMG cluster is characterised by odds ratios below unity for both $f_{\mathrm{gas}}$ and $q_{\mathrm{p/s}}$, indicating that lower values of these parameters increase the likelihood of assignment to this group. The ice--rock mass ratio yields odds ratios close to unity for most clusters.

\begin{table}
\caption{Odds ratios from multinomial logistic regression models for the 3D-mapped synthetic populations.}
\label{tab:logit_odds}
\centering
\small
\renewcommand{\arraystretch}{1.2}
\begin{tabular}{lcccc}
\hline
Parameter & HG (A1) & WJD (A2) & LMG (B) & VMGG (C) \\
\hline
$\mathcal{G}_1$              & 0.74 & 0.57 & 1.33 & 1.79 \\
$f_{\mathrm{gas}}$           & 12.56 & 1.51 & 0.05 & 1.00 \\
Ice--Rock                    & 1.02 & 1.06 & 0.91 & 1.02 \\
$q_{\mathrm{p/s}}$           & 0.02 & 7.58 & 0.04 & 158.38 \\
$a$                          & 0.26 & 16.56 & 0.59 & 0.39 \\
\hline
\end{tabular}
\begin{tablenotes}
\footnotesize
\item Notes: The odds ratios represent the relative change in the probability of assignment to each cluster per $+1\sigma$ increase in the corresponding standardised formation parameter.
\end{tablenotes}
\end{table}

The classification performance of the multinomial logistic regression, used here as a diagnostic tool, is summarised in Table~\ref{tab:logit_performance}. An overall accuracy of 0.982 is achieved on the training set, indicating that the selected formation parameters provide a strong basis for separating the clusters. High precision and recall values are obtained for the LMG and VMGG clusters, while the WJD cluster exhibits lower precision, reflecting its smaller sample size and greater overlap with neighbouring clusters. Both macro-averaged and weighted-average scores remain high, demonstrating that the inferred cluster separation remains robust for both minority and majority components within the synthetic population.

\begin{table}
\caption{Classification performance of the multinomial logistic regression used as a diagnostic tool for the 3D-mapped synthetic populations.}
\label{tab:logit_performance}
\centering
\renewcommand{\arraystretch}{1.2}
\begin{tabular}{lcccc}
\hline
Cluster & Precision & Recall & F1-score & Support \\
\hline
HG (A1)  & 0.912 & 0.959 & 0.935 & 97 \\
WJD (A2) & 0.934 & 0.986 & 0.959 & 72 \\
LMG (B)  & 1.000 & 0.991 & 0.996 & 573 \\
VMGG (C) & 0.990 & 0.950 & 0.970 & 101 \\
\hline
Macro avg    & 0.959 & 0.972 & 0.965 & 843 \\
Weighted avg & 0.983 & 0.982 & 0.982 & 843 \\
\hline
\end{tabular}
\begin{tablenotes}
\footnotesize
\item Notes: An overall accuracy of 0.982 is achieved on the training set; detailed precision, recall, F1-score, and support are listed for each cluster.
\end{tablenotes}
\end{table}

To examine whether formation-related parameters exhibit systematic internal trends within each cluster, rank correlations between selected parameter pairs are evaluated using Kendall’s $\tau$ (Table~\ref{tab:kendall_tau_clusters}). In the HG and WJD clusters, positive correlations are found between $\mathcal{G}_1$ and the ice--rock mass ratio, with the strongest and most statistically significant association occurring in WJD. Correlations involving $f_{\mathrm{gas}}$ in these two clusters are weaker, being negligible in HG and moderate but only marginally significant in WJD.

\begin{table}
\caption{Kendall’s $\tau$ rank correlations between selected formation parameters within each cluster.}
\label{tab:kendall_tau_clusters}
\centering
\renewcommand{\arraystretch}{1.2}
\small
\setlength{\tabcolsep}{2.5pt}
\begin{tabular}{lccc}
\hline
{} 
& $\tau(\mathcal{G}_1,\mathrm{Ice\mbox{--}Rock})$
& $\tau(\mathcal{G}_1,f_{\mathrm{gas}})$
& $\tau(f_{\mathrm{gas}},\mathrm{Ice\mbox{--}Rock})$ \\
\hline
HG (A1) 
& 0.62 ($5.2\times10^{-5}$)
& 0.22 ($1.9\times10^{-1}$)
& 0.28 ($8.6\times10^{-2}$) \\

WJD (A2) 
& 0.86 ($1.7\times10^{-3}$)
& 0.57 ($6.1\times10^{-2}$)
& 0.43 ($1.8\times10^{-1}$) \\

LMG (B)  
& $-$0.50 ($6.1\times10^{-5}$)
& $-$0.42 ($7.8\times10^{-4}$)
& 0.69 ($9.0\times10^{-10}$) \\

VMGG (C)  
& 0.48 ($1.0\times10^{-2}$)
& 0.78 ($2.7\times10^{-5}$)
& 0.25 ($1.9\times10^{-1}$) \\
\hline
\end{tabular}
\begin{tablenotes}
\footnotesize
\item Note. Entries give Kendall’s $\tau$ and the corresponding $p$-value (in parentheses).
\end{tablenotes}
\end{table}

The LMG cluster exhibits weak to moderate negative correlations between $\mathcal{G}_1$ and both the ice--rock mass ratio and $f_{\mathrm{gas}}$, together with a strong positive correlation between $f_{\mathrm{gas}}$ and the ice--rock ratio. For the VMGG cluster, a strong and statistically significant positive correlation is found between $\mathcal{G}_1$ and $f_{\mathrm{gas}}$, together with a moderate correlation between $\mathcal{G}_1$ and the ice--rock mass ratio. Taken together, these results indicate that each cluster is associated with a distinct combination of formation-related properties and parameter correlations.

\end{appendix}

\end{document}